\newif\ifdraft
\newif\ifonecolumn

\draftfalse

\onecolumntrue

\documentclass[10pt,twocolumn]{article}%
\usepackage[utf8]{inputenc}
\usepackage[T1]{fontenc}
\usepackage[english]{babel}
\usepackage{amsmath,amssymb,amsfonts,amsthm}
\usepackage{graphicx}
\usepackage{fancyhdr}
\usepackage{hyperref}
\usepackage{longtable}
\usepackage{listings}
\usepackage{color}
\usepackage{multirow}
\usepackage{wrapfig}
\usepackage{enumitem}
\usepackage{footnote}
\usepackage{tikz,times}
\usepackage{pdflscape}
\usepackage{bbm}
\usepackage{multirow}
\usepackage{comment}
\usepackage{hhline}
\usepackage{graphicx}
\usepackage{caption,subcaption}
\usepackage{cleveref}
\Crefname{equation}{Eq.}{Eqs.}
\Crefname{figure}{Fig.}{Figs.}
\Crefname{tabular}{Tab.}{Tabs.}
\Crefname{table}{Tab.}{Tabs.}

\usepackage{colortbl}
\usepackage{siunitx}
\usepackage{booktabs}
\usepackage{csvsimple}
\usepackage{datatool}
\usepackage{array}
\usepackage{multibib}
\newcites{SM}{References: Supplementary material}
\newcolumntype{L}[1]{>{\raggedright\let\newline\\\arraybackslash\hspace{0pt}}m{#1}}
\newcolumntype{C}[1]{>{\centering\let\newline\\\arraybackslash\hspace{0pt}}m{#1}}
\newcolumntype{R}[1]{>{\raggedleft\let\newline\\\arraybackslash\hspace{0pt}}m{#1}}

\usetikzlibrary{mindmap,backgrounds}

\makesavenoteenv{tabular}
\makesavenoteenv{table}

\definecolor{1color}{HTML}{1F77B4}
\definecolor{2color}{HTML}{FF7F0E}
\definecolor{3color}{HTML}{2CA02C}
\definecolor{4color}{HTML}{D62728}
\definecolor{5color}{HTML}{9467BD}

\definecolor{codegreen}{HTML}{2CA02C}
\definecolor{codegray}{rgb}{0.01,0.01,0.01}
\definecolor{codepurple}{HTML}{1F77B4}
\definecolor{backcolour}{rgb}{0.95,0.95,0.95}

\newcommand\heading[1]{}
\newcommand\xheading[1]{{\noindent\textbf{#1}}}

\lstdefinestyle{mystyle}{
	language=Python,
    backgroundcolor=\color{backcolour},   
    commentstyle=\bfseries\color{codegreen},
    keywordstyle=\bfseries\color{codepurple},
    numberstyle=\tiny\color{codegray},
    stringstyle=\bfseries\color{codegreen},
    basicstyle=\footnotesize\ttfamily,
    breakatwhitespace=false,         
    breaklines=true,                 
    captionpos=b,                    
    keepspaces=true,                 
    numbers=none,                    
    numbersep=5pt,                  
    showspaces=false,                
    showstringspaces=false,
    showtabs=false,                  
    tabsize=2,
    frame=tb
}
 
\usepackage{array}

\usepackage{lineno}
\usepackage[margin=0.75in]{geometry}

\usepackage[affil-it]{authblk} 

\usepackage{wrapfig}

\author[1,*]{Nils Strodthoff}
\author[1]{Juan Miguel Lopez Alcaraz}
\author[2]{Wilhelm Haverkamp}
\affil[1]{Carl von Ossietzky Universität Oldenburg, Oldenburg, Germany\\
     \texttt{\{nils.strodthoff,juan.lopez.alcaraz\}@uol.de}}
\affil[2]{Charit\'e Universit\"atsmedizin Berlin, Berlin, Germany\\
      \texttt{wilhelm.haverkamp@dhzc-charite.de}}
\affil[*]{Corresponding author}

\usepackage{pifont}

\usepackage{comment}

\usepackage{graphicx}      
\usepackage{subcaption}    
\usepackage{lipsum}        

\usepackage{xcolor,cancel}
\usepackage{booktabs}
\usepackage{multirow}
\usepackage{longtable}
\usepackage{enumitem}
\usepackage{booktabs}

\usepackage[normalem]{ulem} 
\newcommand{\stkout}[1]{\ifmmode\text{\sout{\ensuremath{#1}}}\else\sout{#1}\fi}

\UseRawInputEncoding

\begin{document}

\title{Prospects for AI-Enhanced ECG as a Unified Screening Tool for Cardiac and Non-Cardiac Conditions -- An Explorative Study in Emergency Care }

\ifonecolumn
\onecolumn
\else
\twocolumn[
  \begin{@twocolumnfalse}
\fi
    
    \maketitle
    
    Current deep learning algorithms designed for automatic ECG analysis have exhibited notable accuracy. However, akin to traditional electrocardiography, they tend to be narrowly focused and typically address a singular diagnostic condition. In this exploratory study, we specifically investigate the capability of a single model to predict a diverse range of both cardiac and non-cardiac discharge diagnoses based on a sole ECG collected in the emergency department. 
    We find that 253, 81 cardiac, and 172 non-cardiac, ICD codes can be reliably predicted in the sense of exceeding an AUROC score of 0.8 in a statistically significant manner. This underscores the model's proficiency in handling a wide array of cardiac and non-cardiac diagnostic scenarios which demonstrates potential as a screening tool for diverse medical encounters.
    

  \ifonecolumn
  \else
  \hspace{2ex}  
  \end{@twocolumnfalse}
]
\fi

\section{Introduction}
\heading{AI-enhanced ECG} 
The electrocardiogram (ECG) holds a distinctive role as the primary tool for assessing a patient's cardiac status, with over one-fourth of U.S. emergency department visits involving an ECG \cite{NHAMCS2021}. Presently, manual assessment predominates, with limited algorithmic support from rule-based ECG devices, known for their constraints \cite{schlapfer2017computer}. The emergence of deep learning, has sparked interest in AI-enhanced ECG interpretation, revolutionizing diagnostic perspectives \cite{TOPOL2021785, Siontis2021}. Numerous studies showcase deep learning's accuracy in inferring diverse cardiac conditions, from myocardial infarction and comprehensive ECG statements \cite{Strodthoff:2020Deep,Kashou2020} to rhythm abnormalities \cite{Hannun2019}. Remarkably, deep learning models demonstrate proficiency in inferring age, sex \cite{Attia2019}, ejection fraction \cite{Attia2019b}, atrial fibrillation during sinus rhythm \cite{attia2019artificial}, anemia \cite{Kwon2020}, and non-cardiac conditions like diabetes \cite{kulkarni2023machine} and cirrhosis \cite{Ahn2021}, challenging for human experts to discern from an ECG.

\begin{figure*}[!ht]
    \centering
    \includegraphics[width=0.9\textwidth]{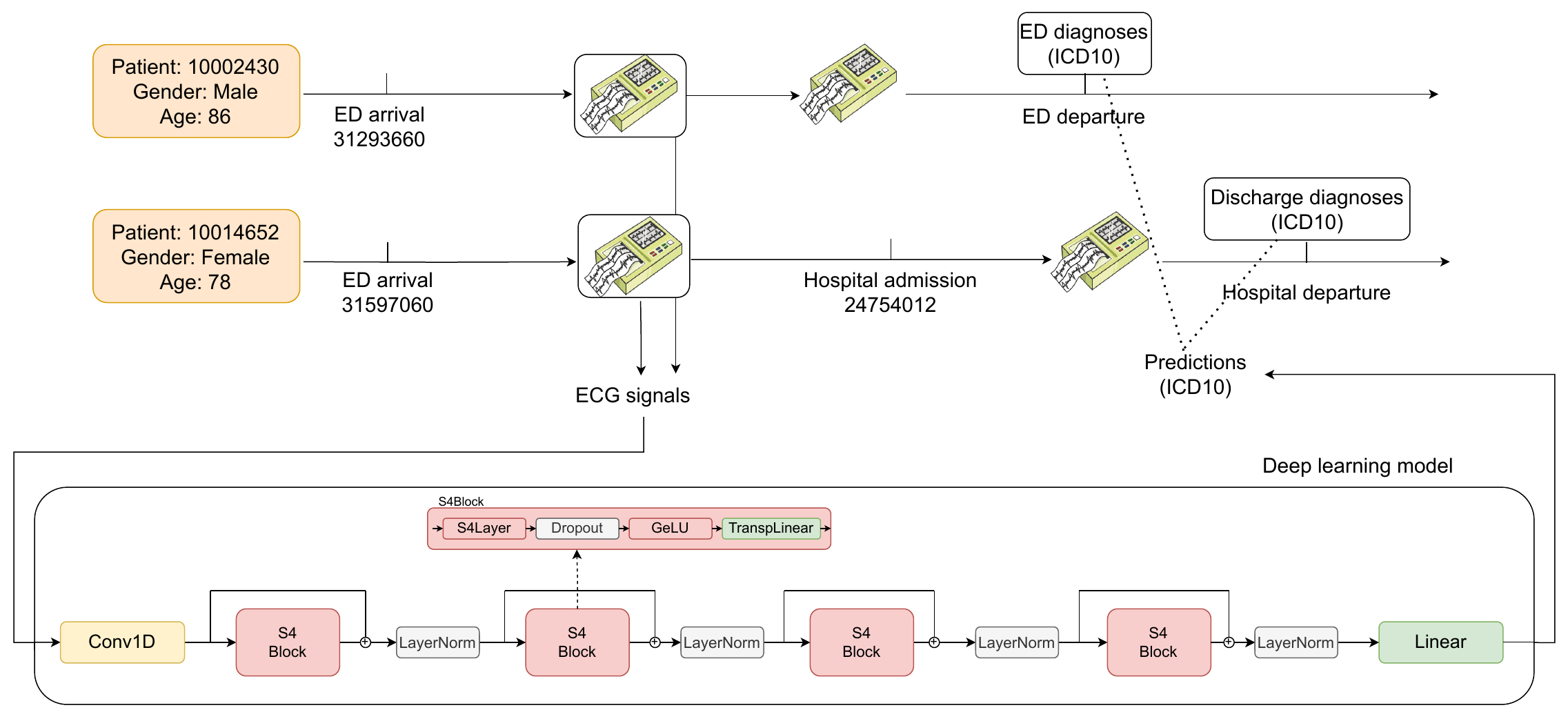}
    \caption{Schematic illustration depicting the proposed workflow, featuring two patient use cases. First, consider Patient 10002430, who does not undergo hospital admission, we try to infer the ED discharge diagnosis from the initial 10s of ECG. This snippet is fed into our deep learning model, which outputs probabilities for each of the most reliably predictable ICD-10 codes (e.g., 439 codes with AUROC exceeding 0.8). Second, consider the arrival of Patient 10014652 to the Emergency Department (ED), where a variety of ECG recordings are obtained during both the ED stay and subsequent hospital admission. In this particular scenario, our objective is to predict the hospital discharge diagnosis, again based on the first 10 seconds of the recorded ECG. This approach allows us to leverage the most accurate clinical ground truth and to connect it to the first recorded ECG of the patient, which can provide valuable information for decisions at the ED.
    }
    \label{fig: pipeline}
\end{figure*}

\heading{Limitations of existing work}
While notable AI-enabled ECG studies demonstrate impressive performance, a prevalent limitation is their narrow scope. Typically confined to binary prediction problems, these studies face challenges in defining appropriate control groups, potentially leading to an overestimation of algorithmic performance in real-world scenarios. Additionally, these studies are almost exclusively based on closed-source datasets, which hinder reproducibility and scientific progress. The availability of public ECG datasets has increased considerably \cite{Wagner:2020PTBXL}, however, they typically lack clinical ground truth, limiting their utility for uncovering the diagnostic boundaries of the ECG. Finally, the emergence of specialized FDA-approved ECG algorithms raises questions about the feasibility of numerous isolated apps with limited scope, overlooking the intricate clinical reality of co-occurring diseases.

\heading{Related work} Existing works violate at least one of the points raised above. First of all, there is no comprehensive prediction algorithm beyond cardiovascular conditions based on raw ECGs as input. Even for cardiovascular diseases, binary conditions are the most common setup, with a few notable exceptions. \cite{Ribeiro2020} cover different cardiovascular conditions, but restrict themselves to a rather coarse set of 6 conditions. \cite{Kashou2020} achieve excellent results for the prediction of 66 cardiovascular conditions, which still fall short compared to the more than 150 cardiovascular conditions considered in this work, and base their work exclusively on a closed in-hospital dataset. Prediction models trained on public ECG datasets \cite{Wagner:2020PTBXL} such as \cite{Strodthoff:2020Deep} cover a somewhat extensive set of cardiovascular conditions, but lack clinical ground truth for more comprehensive investigations. Finally, \cite{sun2022ecg} is closest to our work as they also address discharge diagnosis prediction from the raw ECG, however, exclusively work on a closed in-hospital dataset and do not provide any external validation.

\heading{Vision of a holistic ECG analysis}
As already mentioned above, many, not exclusively, cardiac conditions leave traces in the ECG. However, apart from a small, selected number of conditions, this question has not been answered comprehensively, see \cite{Kashou2022} for a recent perspective. We envision that a deep-learning-based ECG analysis algorithm trained on a comprehensive set of a general set of clinical diagnostic statements could provide patient profiles with detailed personalized risk scores (after appropriate calibration). Furthermore, learned features of such models could be used for deep phenotyping, in this case, obtained from supervised fine-tuning very much analogous to the widely used models pre-trained on ImageNet in computer vision, complementary to advances in self-supervised pre-training \cite{Mehari:2021Self}. These offer exciting prospects in terms of patient retrieval, also in combination with patient profiles from other modalities such as whole-genome sequencing or foundation models for medical imaging.

\heading{Use-case: Triage in the emergency department}
As a demonstration, we consider a subset of the full dataset under consideration of ECGs that were taken at the emergency department (ED) and investigate the feasibility of predicting ED diagnoses or (if available) hospital discharge diagnoses from them. The specific use case we have in mind is the triage in the ED, where electronic differential diagnostic support could lead to a significant reduction in diagnostic errors if integrated properly into the scope and the context of the ED triage process \cite{Sibbald2022}. The proposed model could be further supplemented by basic patient metadata such as patient demographics, chief complaints, and basic lab values to further improve model accuracy and robustness.

\heading{Contributions}
In this exploratory study, we address the above limitations and explore the potential of deep learning in predicting a broad range of diagnoses, i.e., cardiac and non-cardiac discharge diagnoses from a single 12-lead ECG, with an application as a screening method in an ED setting in mind, based exclusively on publicly available data. We construct the \textit{MIMIC-IV-ECG-ICD-ED} dataset from publicly available MIMIC-IV-ECG and MIMIC-IV data. State-of-the-art prediction models are trained and evaluated on this dataset, showing strong performance across over 1000 cardiac and non-cardiac conditions. We perform an initial external validation and compare our model's performance to narrower-scope prediction models from the literature, discussing implications for ED triage.

\section{Methods}

\subsection{Dataset construction and preprocessing}

The proposed \textit{MIMIC-IV-ECG-ICD(-ED)} dataset was created by linking signals from the MIMIC-IV-ECG \cite{MIMICIVECG2023} dataset to clinical ground truth from the clinical MIMIC-IV dataset \cite{Johnson2023}. This involved aligning ECG recording times with patient admission and discharge times, retrieval and standardization of diagnostic codes (ICD-9-CM to ICD-10-CM), where hospital discharge diagnoses was given preference over ED diagnoses due to higher comprehensiveness and reliability. A detailed description of the dataset construction and preprocessing steps can be found in the supplementary material.

\subsection{Prediction tasks and training procedures}

The prediction task is a multi-label classification, where each patient's discharge diagnosis is a set of ICD-10 statements, capturing clinical complexity comprehensively. We use all ECGs in the training set and minimize binary cross-entropy loss for multi-label prediction. Models are optimized using AdamW with a learning rate of 0.001 and weight decay of 0.001, trained for 20 epochs with a batch size of 32. We applied model selection based on the highest macro AUROC on the validation set to prevent overfitting, where the best-perfoming model was typically found around epoch 15. Prior research \cite{Strodthoff:2020Deep} showed better performance by averaging predictions from shorter 2.5s crops, despite models' ability to appropriately handle long-range interactions \cite{Mehari2023S4}. Therefore, we train on random 2.5s crops and average predictions over four non-overlapping crops during testing. A single model training took approximately 19 hours on a single NVIDIA A30 GPU.

\subsection{Evaluation procedures}

In contrast to the model training process, the test and validation sets only include the first ECG per ED/hospital stay per patient to prevent bias in model evaluation from patients with a large number of ECGs per stay. The primary evaluation metric is the macro average across all areas under the respective receiver operating curves (AUROC) (macro AUROC). To assess statistical uncertainty resulting from the finite size and specific composition of the test set, we employ empirical bootstrap on the test set with $n=1000$ iterations. We report 95\% confidence intervals for both macro AUROC and individual label AUROCs.

We primarily focus on the ED use case in our proposed dataset, enabling investigation into various conditions based on subsets used for training/evaluation and label sets, which may not necessarily coincide. To differentiate between them, we introduce the notation T($A$2$B$)-E($C$2$D$), where $A,C\in{\text{ALL},\text{ED},\text{HOSP}}$ refers to the subset of ECGs used for training/evaluation and $B,D\in{\text{ALL},\text{ED},\text{HOSP}}$ refers to the label sets used for training/evaluation. The main scenario is denoted as T(ED2ALL)-E(ED2ALL). In the supplementary material, we compare this model with one trained on the most comprehensive dataset, T(ALL2ALL)-E(ALL2ALL), and explore various cross-evaluation scenarios, such as evaluating the comprehensive model on the ED subset (T(ALL2ALL)-E(ED2ALL)). Notably, the model trained on the most comprehensive dataset, T(ALL2ALL)-E(ED2ALL), achieved slightly lower performance compared to the specialized T(ED2ALL)-E(ED2ALL) model, with macro AUCs of 0.7691 and 0.7742, respectively, which was statistically significant. However, the specialized model performs considerably weaker across different evaluation scenarios. Detailed descriptions of these scenarios and extensive performance comparisons are provided in the supplementary material.

\begin{figure*}[!ht]
    \centering
    \includegraphics[width=0.9\textwidth]{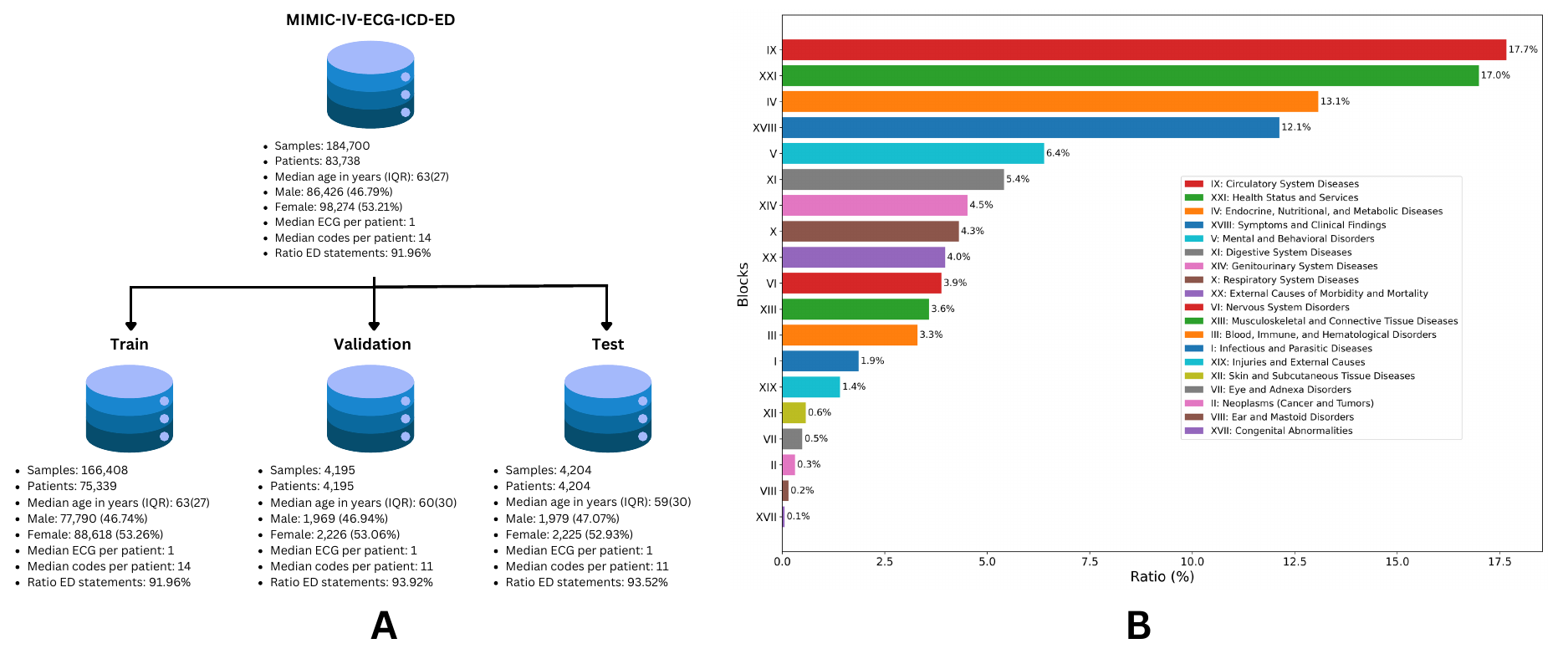}
    \caption{Schematic summary of the dataset composition and distribution of ICD codes across the dataset. (A) From the main MIMIC-IV-ECG-ICD-ED database of 184,700 samples across 83,738 patients, we utilize records of 75,339 patients for training, records of 4,195 patients for model selection in the validation stage, and records of 4,204 patients for testing. The median ECG records per patient is 1, however, the distribution is long-tailed with a maximum of ECG records per patient at 171.
    (B) represents the distribution of ICD codes according to chapters (all percentages as relative fractions compared to the dataset size), where chapter IX (Circulatory system diseases) is the most strongly represented chapter with 17.7\%, closely followed by chapter XXI (Health system and status) with 17\%, we present in supplementary material the distribution of cardiac conditions within chapter IX (Circulatory system diseases categories).}
    \label{fig: summary}
\end{figure*}

\section{Results}

\subsection{MIMIC-IV-ECG-ICD-ED Dataset}

We construct an ECG dataset with clinical labels, \textit{MIMIC-IV-ECG-ICD}, as a subset of the MIMIC-IV-ECG dataset \cite{MIMICIVECG2023}  obtained by joining its records with hospital discharge diagnosis or ED diagnosis (in case the former is unavailable) from the MIMIC-IV dataset \cite{Johnson2023}. In this work, we will only work with the subset of ECGs captured in the ED and refer to it as \textit{MIMIC-IV-ECG-ICD-ED} dataset.
In \Cref{fig: summary}, (A) we illustrate the dataset composition along with corresponding descriptive statistics, similarly, in (B) we summarize the ED subset in terms of the label distribution according to ICD-10 chapters.

\subsection{Model performance}
\heading{Global performance} We report classification results for a structured state-space sequence (S4) model \cite{gu2021efficiently}, which outperformed state-of-the-art convolutional model, see the additional results in the supplementary material, confirming earlier findings \cite{Mehari2023S4}. In the first column in \Cref{tab:big_all}, we summarize performance according to ICD-10 chapters with the most predictive chapters IX (Circulatory system diseases; AUROC 0.841) and X (Respiratory system diseases; AUROC 0.803). In 439 out of the 1079 considered ICD-10 codes the performance exceeds an AUROC score of 0.8. The statistical uncertainty assessed via bootstrap confidence intervals is reasonably low with a median of 0.0506 (IQR 0.0373) across all codes. As a more conservative criterion for discovery, we consider only those codes for which the lower bound of the bootstrap confidence interval exceeds 0.8, i.e., conditions where the performance exceeds 0.8 in a statically significant manner. This singles out 253 ICD codes, 81 of which are cardiac and 172 of which are non-cardiac.

\begin{table*}[ht!]
\centering
\scalebox{0.60}{
\begin{tabular}{lp{10cm}p{10cm}}
\toprule
Block: Block description. Block AUROC & Code: Code AUROC. Code description &  Code: Code AUROC. Code description \\  
\midrule

\midrule
\multirow{15}{*}{IX: Circulatory System Diseases. 0.843} & I210: 0.986. ST elevation (STEMI) myocardial infarction of anterior wall & I314: 0.979. Cardiac tamponade\\
&I447: 0.976. Left bundle-branch block, unspecified & I481: 0.966. Persistent atrial fibrillation\\
&I451: 0.964. Other and unspecified right bundle-branch block & I255: 0.964. Ischemic cardiomyopathy\\
&I132: 0.948. Hypertensive heart and chronic kidney disease with heart failure and with stage 5 chronic kidney disease, or end stage renal disease & I078: 0.948. Other rheumatic tricuspid valve diseases\\
&I081: 0.944. Rheumatic disorders of both mitral and tricuspid valves & I2789: 0.943. Other specified pulmonary heart diseases\\
&I5043: 0.94. Acute on chronic combined systolic (congestive) and diastolic (congestive) heart failure & I7025: 0.927. Atherosclerosis of native arteries of other extremities with ulceration\\
&I340: 0.913. Nonrheumatic mitral (valve) insufficiency & I428: 0.906. Other cardiomyopathies\\
&I110: 0.906. Hypertensive heart disease with heart failure & I359: 0.897. Nonrheumatic aortic valve disorder, unspecified\\
&I851: 0.895. Secondary esophageal varices & I200: 0.891. Unstable angina\\
&I46: 0.89. Cardiac arrest & I120: 0.888. Hypertensive chronic kidney disease with stage 5 chronic kidney disease or end stage renal disease\\

\midrule    
\multirow{2}{*}{X: Respiratory System Diseases. 0.804} & J9621: 0.951. Acute and chronic respiratory failure with hypoxia & J948: 0.925. Other specified pleural conditions\\
&J80: 0.905. Acute respiratory distress syndrome & J910: 0.904. Malignant pleural effusion\\

\midrule
\multirow{2}{*}{II: Neoplasms (Cancer and Tumors). 0.798} & C7952: 0.933. Secondary malignant neoplasm of bone marrow & C925: 0.922. Acute myelomonocytic leukemia\\
&C25: 0.884. Malignant neoplasm of pancreas & D469: 0.869. Myelodysplastic syndrome, unspecified\\

\midrule
\multirow{3}{*}{XXI: Health Status and Services. 0.793} & Z998: 0.946. Dependence on other enabling machines and devices & Z681: 0.944. Body mass index (BMI) 19.9 or less, adult\\
&Z4502: 0.941. Encounter for adjustment and management of automatic implantable cardiac defibrillator & Z950: 0.94. Presence of cardiac pacemaker\\

\midrule
\multirow{3}{*}{XII: Skin and Subcutaneous Tissue Diseases. 0.790} & L891: 0.875. Pressure ulcer of back & L9740: 0.87. Non-pressure chronic ulcer of unspecified heel and midfoot\\
&L0312: 0.851. Acute lymphangitis of other parts of limb & \\

\midrule
\multirow{3}{*}{IV: Endocrine, Nutritional, and Metabolic Diseases. 0.785}& E1129: 0.924. Type 2 diabetes mellitus with other diabetic kidney complication & E660: 0.907. Obesity due to excess calories\\
&E103: 0.899. Type 1 diabetes mellitus with ophthalmic complications & E43: 0.886. Unspecified severe protein-calorie malnutrition\\

\midrule
\multirow{4}{*}{XIX: Injuries and External Causes. 0.777} &  T8612: 0.944. Kidney transplant failure & T8285: 0.898. Stenosis due to cardiac and vascular prosthetic devices, implants and grafts\\
&T380: 0.898. Poisoning by, adverse effect of and underdosing of glucocorticoids and synthetic analogues &  \\

\midrule
\multirow{3}{*}{I: Infectious and Parasitic Diseases. 0.771} & B9620: 0.862. Unspecified Escherichia coli [E. coli] as the cause of diseases classified elsewhere & A419: 0.86. Sepsis, unspecified organism\\
&A40: 0.859. Streptococcal sepsis & \\

\midrule
\multirow{2}{*}{V: Mental and Behavioral Disorders. 0.766} & F1022: 0.894. Alcohol dependence with intoxication & F1721: 0.88. Nicotine dependence, cigarettes\\
&F4310: 0.864. Post-traumatic stress disorder, unspecified & \\

\midrule
\multirow{3}{*}{XIV: Genitourinary System Diseases. 0.759} & N186: 0.887. End stage renal disease & N08: 0.878. Glomerular disorders in diseases classified elsewhere\\
&N9982: 0.857. Postprocedural hemorrhage of a genitourinary system organ or structure following a procedure & N170: 0.852. Acute kidney failure with tubular necrosis\\

\midrule
\multirow{3}{*}{III: Blood, Immune, and Hematological Disorders. 0.755}& D65: 0.933. Disseminated intravascular coagulation [defibrination syndrome] & D684: 0.878. Acquired coagulation factor deficiency\\
& D631: 0.857. Anemia in chronic kidney disease & \\

\midrule
\multirow{2}{*}{XI: Digestive System Diseases. 0.741} & K7031: 0.973. Alcoholic cirrhosis of liver with ascites & K762: 0.948. Central hemorrhagic necrosis of liver\\
&K7290: 0.947. Hepatic failure, unspecified without coma & K3189: 0.921. Other diseases of stomach and duodenum\\

\midrule
\multirow{2}{*}{XVIII: Symptoms and Clinical Findings. 0.7162} &  R570: 0.931. Cardiogenic shock & R64: 0.9. Cachexia\\
&R18: 0.887. Ascites & R6521: 0.887. Severe sepsis with septic shock\\

\bottomrule
\end{tabular}
}

\caption{Best-performing individual statements organized according to selected ICD chapters underscoring the breadth of accurately predictable statements. The table shows the four best-performing individual statements per ICD chapter (20 for chapter IX (Circulatory system diseases)), where we show only AUROC point predictions above 0.85 where also the lower bound of the 95\% bootstrap confidence interval exceeds 0.80. To showcase the breadth of reliably predictable statements, we list only the best-performing statement per 3-digit ICD code. The complete list of AUROC scores for all 1076 ICD codes is provided in the supplementary material as a summary of ICD codes at a 3-digit level with AUROC scores above 0.9, 0.8, and below 0.7 respectively.}

\label{tab:big_all}
\end{table*}

\heading{Most accurately predictable individual statements}
\Cref{tab:big_all} gives a comprehensive overview of conditions grouped by ICD chapters based on their predictability measured by AUROC scores. This list provides a broad view that can be seen as an exploratory approach toward many clinical directions. It contains a wide range set of cardiac conditions such as ST-elevation myocardial infarction, cardiac tamponade, left/right bundle branch block, persistent atrial fibrillation, or ischemic cardiomyopathy (all with AUROC scores above 0.95). Notably, there were also non-cardiac codes with exceptionally high predictive performance such as E. coli, leukemia, type 2 diabetes, alcohol dependence, respiratory failure, alcoholic liver cirrhosis, ulcer, renal disease, cardiogenic shock, poisoning, traffic accidents, or the presence of assistance devices.

\heading{Very accurately predictable ICD codes (AUROC > 0.9)}
We analyze the results underlying \Cref{tab:big_all} from a complementary perspective by aggregating codes on the level of 3-digit ICD codes and indicating also the \textit{coverage}, i.e., the fraction of sub-statements (including the 3-digit code itself) of a particular 3-digit code that exceeds a predefined accuracy threshold (in this case chosen as AUROC > 0.9). A high coverage indicates that the model has acquired a good understanding of the particular condition including its corresponding differential diagnoses. Focusing on statements with a coverage of 75\% or more, we see that the ECG is highly predictive for a wide range of cardiac conditions such as atrial fibrillation, hypertensive heart diseases, left bundle branch block, acute myocardial infarction, and heart failure. Notable non-cardiac conditions include pleural conditions, (alcoholic) liver diseases, traffic accidents, and assistant-device-related conditions.

\heading{Accurately predictable ICD codes (AUROC > 0.8): Cardiac conditions}
We again assess groups of statements that can be reliably predicted from the ECG, however, this time at a slightly lower accuracy threshold of AUROC scores above 0.8. In the discussion below we focus again on statements with a high (75\%) coverage. Cardiovascular conditions from Chapter IX most notably include chronic ischemic heart diseases, atrial fibrillation, heart failure, hypertension, pulmonary heart diseases, acute myocardial infarction, valve disorders, cardiomyopathy, left bundle branch blocks, and other conduction disorders.

\heading{Accurately predictable ICD codes (AUROC > 0.8): Non-cardiovascular conditions} 
Our results imply that the ECG may have predictive capabilities across a very broad range of conditions including sepsis (I), neoplasms and leukemia (II), anemia and disseminated intravascular coagulation (III), diabetes type I, overweight and malnutrition, and partly also diabetes type II (IV), dementia and psychoactive drugs (V), Alzheimer's and Parkinson's (VI), respiratory failure, pleural effusion (X), liver diseases, alcoholic liver, stomach diseases, hepatic failure from (XI), ulcer (XII), gout (XIII), benign prostatic hyperplasia and chronic kidney diseases (XIV), heart valve malformation (XVII), systemic inflammation, shock (XVIII), different kinds of poisoning (XIX), traffic accidents (XX), presence of cardiac implants or assistance devices, body mass index, absence of limb, and implanted device management (XXI). In parentheses, we always indicate the respective ICD chapter, which is also listed in full form in \Cref{tab:big_all}.

\begin{table}[!ht]
    \centering
    \begin{tabular}{cccc}
    \toprule
    Statement & ICD-10 codes & Internal & External \\
    \midrule
    1AVB & I440 & 0.908 & 0.942 \\
    AFIB & I4891 & 0.908 & 0.970 \\
    LBBB & I447 &  0.976 &  0.999 \\
    RBBB & I4510 & 0.964 &  0.989 \\
    SBRAD & R001 & 0.791 &  0.957 \\
    STACH & R000 & 0.849 &  0.985 \\
    \bottomrule
    \end{tabular}
    \caption{External validation on CODE-test for diverse cardiac conditions (1AVB: 1st degree AV block, AFIB: atrial fibriallation, LBBB: left bundle branch block, RBBB: right bundle branch block, SBRAD: sinus bradycardia, STACH: sinus tachycardia). We report AUROC scores for specified ICD-10 codes on the internal and on the external  CODE-test dataset.}
    \label{tab:external}
\end{table}

\heading{External validation}

We acknowledge the importance of an external validation \cite{Chekroud2024} and therefore validate the performance of our model on the CODE test set \cite{Ribeiro2020} across a set of six different cardiac conditions. The mapping between conditions and ICD-10 codes is non-trivial, so we picked the most typical ICD-10 code that aligns with it. Across all conditions in \Cref{tab:external}, the model exhibits an even stronger performance than on the internal test set despite a sizeable mismatch in label distribution across both datasets. As the most likely explanation, it is worth stressing that the prediction task on the internal is much more fine-grained requiring to differentiate between similar differential diagnoses instead of differentiating within a coarse set of only 6 conditions. The comparably weak performance on the bradycardia and tachycardia conditions might also be affected by missing annotations in the internal dataset, which are not ECG-specific annotations and only include the most important persistent conditions.
There is presently no publicly available ECG dataset covering an ED patient cohort, which would qualify as an external validation dataset covering non-cardiac conditions.

\section{Discussion}

\subsection{AI-enhanced ECG as a unified screening tool}

\heading{Introduction and benchmarking}
Our study demonstrates that using deep learning on a single 12-lead ECG effectively predicts both cardiac and non-cardiac conditions for discharge diagnoses, making it a valuable screening tool in an Emergency Department (ED) setting. In line with the explorative nature of this investigation, we see a large number of accurately predictable (also non-cardiac) statements as a strong hint at the diagnostic power of the AI-enhanced ECG, which remains to be validated in detailed follow-up studies. In addition to the external validation for common cardiac conditions, we further validate our model by comparing its performance to existing predictive models from the literature, which, however suffers from systematic uncertainties due to varying definitions of conditions and limited coverage of relevant pathologies in control groups. To address these challenges, the proposed \textit{MIMIC-IV-ECG-ICD} dataset facilitates standardized comparisons with clinical ground truth, similar to PTB-XL \cite{Wagner:2020PTBXL}, to accelerate progress in the field.

\heading{Comparison to landmark results}
In addition to our external validation, we set our results into perspective by comparing them to landmark results from recent literature. \cite{Attia2019b} report an AUROC score of 0.932 for the detection of a low ejection fraction of less than 35\%, which is often associated with heart failure. To put this into perspective, for \textit{heart failure with reduced ejection fraction} we report an AUROC of 0.936. Another landmark paper assessed AF from sinus rhythm with an AUROC of 0.87. We compare this to the performance of our model on \textit{paroxysmal atrial fibrillation}, which reaches an AUROC of 0.891. Turning to non-cardiac conditions, \cite{ahn2022development} have developed predictive models for the detection of cirrhosis from ECGs with an AUROC of 0.908, we report 0.906 AUROC for \textit{cirrhosis} detection as well as 0.973 for \textit{cirrhosis with ascites}. Finally, \cite{Kwon2020} demonstrated the feasibility of predicting \textit{anemia} from the ECG, reporting an AUROC score 0.923. We report AUROCs up to 0.857 for different sorts of anemias.

While the above literature comparison might seem very selective, we present an extensive comparison of literature results in the supplementary material. These findings highlight the competitiveness of the proposed model in both cardiac and non-cardiac conditions. A unique strength is its fine-grained predictions. While qualitative ECG changes are known for many non-cardiac conditions, our study provides the first quantitative evidence for their predictability. The alignment with literature results and correspondence with known qualitative ECG changes validate our approach, extending to conditions where predictability is reported for the first time. 

Our model excels in predicting non-cardiac conditions, which 
may appear surprising given their apparent lack of correlation with the heart's electrical function. While commonly thought to solely depict the heart's electrical activity, the ECG is, in fact, intricately influenced by factors such as the autonomic nervous system, gender, hormones, age, weight, and extracardiac elements (thoracic configuration and impedance). Inflammatory or autoimmune diseases \cite{Capecchi2019} or even chest trauma may also cause ECG changes. AI's recognition of certain profiles might be rooted in these extracardiac factors.

\subsection{Limitations}

\heading{Bias in coding and label quality uncertainty}
The proposed approach has several limitations. First, discharge diagnoses may include events unrelated to the patient's condition captured by the ED ECG, and the coding process itself is prone to biases. The former could be addressed by incorporating temporal metadata, whereas the latter could be mitigated through the use of full-text discharge reports. However, it is worth stressing that the discharge diagnoses serve as a proxy for the clinical ground truth and hence represent a qualitative improvement over labels from expert annotations.

As a second limitation, we stress that our findings are associative and do not indicate causal relationships. Confounding factors like demographic variables, concurrent ailments, treatments, and nuanced medical history may obscure causal links. Clinical metadata, including chief complaint summaries, could uncover more intricate factors. 
Thirdly, in addressing co-occurring diseases as confounding factors, our approach is less susceptible to uncontrolled confounding effects compared to common binary approaches. Unlike methods requiring a distinct control group, our utilization of all ED ECGs encompasses a clinically relevant patient collective. Patients without the specific condition implicitly serve as a control group. For instance, a recent study \cite{OuyangMICCAI23} noted confounding in detecting cirrhosis from an ECG due to ascites. Our model explicitly resolves this, achieving high AUROC scores for cirrhosis with and without ascites without creating specific control sets.

\heading{Analysis of label-label correlations}

Lastly, despite efforts to control confounding factors, our approach may still be affected by comorbidities. We analyze correlations within the test set labels using Matthew's correlation coefficients (MCCs) (detailed in supplementary material). Among the 1076 most prevalent labels, the highest correlations exist between specific and parent statements, suggesting that the model might potentially not be able to capture the parent statement in its full breadth. Excluding parent statements reveals correlations across various labels, typically representing variations of codes for the same underlying condition or statements relating to a specific condition and a corresponding treatment (e.g. renal diseases and dialysis with MCC 0.84). Notably, the correlation between type 1 diabetes mellitus with opthalmological and neurological complications stands out with an MCC of 0.64). While our analysis does not strongly suggest significant confounding effects from co-occurring labels, they might have been exploited by the model in selected cases compromising the model's ability for generalization and therefore warrants consideration in future investigations. 

\heading{Analysis of label-demographics correlations}
Using a similar methodology, we also investigated correlations of demographic subgroups (gender and subgroups of patients exceeding a certain age) and all considered diagnostic conditions in the test set.  We identified mostly age-related correlations with certain diagnostic conditions (lipidemia, atrial fibrillation, heart failure, atherosclerotic heart disease, dementia, chronic kidney diseases) all of them with moderate MCCs between 0.2 and 0.3, which could be at least exploited by the model, which could be analyzed using concept-based XAI methods \cite{wagner2023explaining}. However, it does not compromise the study's aim of assessing the predictability of diagnostic statements from the ECG alone but rather provides supporting evidence for the inclusion of additional clinical metadata.

\subsection{Future research directions}
As a promising direction for future work, leveraging explainable AI methods \cite{wagner2023explaining,Vielhaben:2022MCD} could enhance comprehension of disease-related ECG changes as insights from the model's predictions. Similarly,  while our model exclusively uses ECG data, future enhancements should prioritize the inclusion of additional inputs, such as demographic \cite{Mehari2023S4}, chief complaint summaries \cite{Zhou2023}, and basic lab values.

\textbf{Contributors}
NS and WH conceptualized the study. NS produced the first code prototype. JMLA. carried out the full experiments. NS and JMLA summarized the outputs and produced display items.  All authors interpreted the results. NS and JMLA wrote the first draft and all authors revised it. All authors approved the submitted version.

\section*{Data sharing}
This study is based on the publicly available MIMIC-IV-ECG dataset
\cite{MIMICIVECG2023}(\url{https://doi.org/10.13026/4nqg-sb35}) in combination with clinical ground truth from the clinical MIMIC-IV dataset \cite{Johnson2023}(\url{https://doi.org/10.13026/6mm1-ek67}). External validation was carried out based on the CODE test set \cite{Ribeiro2020}.

The source code underlying our investigations is available under \url{https://github.com/AI4HealthUOL/ECG-MIMIC}.

\section*{Declaration of Interests}
The authors declare no competing interests.

\bibliography{bibfile}
\bibliographystyle{ieeetr}

\vfill

\clearpage
\onecolumn
\appendix

\setcounter{page}{1}
\renewcommand\thefigure{A.\arabic{figure}}  
\renewcommand\thetable{A.\arabic{table}}  
\setcounter{figure}{0}  
\setcounter{table}{0}

\section{Supplementary material for ``AI-Enhanced ECG: A Unified Screening Tool for Cardiac and Non-Cardiac Conditions -- An Explorative Study in Emergency Care''}

\subsection{Dataset construction and preprocessing}

To link samples from the MIMIC-IV-ECG dataset \cite{MIMICIVECG2023} to clinical ground truth from the clinical MIMIC-IV dataset \cite{Johnson2023}, we identified ECGs taken in the ED or hospital by comparing recording times to patient admission, discharge, and potential death times. This process yielded ED stay IDs (\textit{stayid}) for ED-captured ECGs and hospital admission IDs (\textit{hadmid}) for ECGs taken in the hospital or for ED ECGs of patients subsequently admitted. ED stay IDs enabled retrieval of ED discharge diagnoses (max. 9 ICD-9-CM or ICD-10-CM codes), while hospital admission IDs link samples to hospital discharge diagnoses (max. 39 ICD-9-CM or ICD-10-CM codes).

We used the Python package \textit{icd-mappings} to convert ICD-9-CM to ICD-10-CM codes, establishing a common vocabulary. All codes were truncated to a consistent five-digit format (e.g., I48.92 for unspecified atrial flutter), retaining entries with fewer digits and removing trailing 'X' placeholder characters. Superclasses up to a three-digit level were included to ensure consistent mapping (e.g., I48.9 and I48 for unspecified atrial fibrillation and atrial flutter, respectively). ECG samples were resampled to 100Hz \cite{Mehari2023S4}, with missing signal values linearly interpolated and infrequent missing values at sequence boundaries replaced with zero. Signals were clipped to a maximum amplitude of 3~mV. Apart from resampling, handling missing values, and clipping, no further preprocessing was applied to the raw ECG signals. The MIMIC-IV-ECG dataset patients were randomly assigned to twenty folds: 18 for training, 1 for validation and model selection, and 1 for testing. Corresponding fold assignments are included in the code repository for reproducibility.

The linking process defines three label sets: ED discharge diagnoses, hospital discharge diagnoses (from ED ECGs of subsequently admitted patients or from hospital ECGs), and a combination of both. In cases of both present, the hospital diagnosis is prioritized for its comprehensiveness (max. 39 vs. 9 ED discharge diagnoses) and higher diagnostic precision. Although training generally occurs on the combined set, various subsets with different label sets are evaluated. A negligible fraction of ECGs with an empty set of discharge diagnoses is discarded, as examination of corresponding discharge reports showed many lacked ICD annotations. The final label sets are selected by discarding codes occurring less than 2000 times in the dataset, resulting in a label set comprising 1076 3- to 5-digit ICD-10-CM codes for training.

\subsection{Literature comparison: overview}
\xheading{Most accurately predictable cardiac codes}
We begin our discussion on cardiovascular conditions with high-performing statements from \Cref{tab:big_all}. Interestingly, several of these statements have primarily been inferred from imaging modalities rather than directly from the ECG. \citeSM{bansal2022machine} predict \textit{cardiac tamponade} after AF diagnoses based on tabular features (patient metadata, previous diagnoses, laboratory values) with 0.84 AUROC. We predict it from a single ECG alone with a significantly higher AUROC score of 0.979. Prior work on \textit{ischemic cardiomyopathy} was typically based on imaging modalities such as echocardiography \citeSM{HAMADA2016412,zhou2023echocardiography} with AUROC scores of 0.934. We report a performance of 0.964 for ischemic cardiomyopathy from the ECG alone. \citeSM{kwon2020artificial} detected \textit{cardiac arrest} (within 24h hours) from ECG signals with a performance of 0.913 AUROC. We report an AUROC of 0.890 without any constraints on the time frame. The assessment of \textit{pulmonary heart diseases} is challenging as the typical signs may be insensitive and often appear late in the disease. We report an AUROC score of 0.943.

\xheading{Accurately predictable cardiac codes}
The list of accurately predictable cardiac conditions is extensive and we provide selected insights prioritized by prevalence. For a more comprehensive discussion, please refer to the supplementary material.
In the domain of \textit{chronic ischemic heart diseases}, \citeSM{ambale2017cardiovascular} used imaging noninvasive tests, and biomarker panels for the detection of atherosclerotic.  We report AUROCs of 0.827 for chronic ischemic heart disease as a broad category, as well as 0.825, and 0.903 for atherosclerotic with native coronary artery without angina pectoris and with angina pectoris, respectively. \citeSM{Sehrawat2022} review deep-learning methods for \textit{atrial fibrillation} and flutter detection by ECGs, which includes the work of \citeSM{kashou2020comprehensive}, with AUROCs ranging from 0.82 to 0.99. Deep learning models \cite{Strodthoff:2020Deep} on open source datasets reach AUROC scores of 0.982. This should be compared to a score of 0.966 for persistent atrial fibrillation. In a more fine-grained setting \citeSM{amaya2022development} reported an AUROC of 0.876 for \textit{paroxysmal atrial fibrillation}, in comparison to 0.891 from our model. \citeSM{hussain2020detecting} used temporal, spectral, and complex HRV dynamics to detect \textit{heart failure} with an AUROC of 0.97. \citeSM{doi:10.4070/kcj.2018.0446} used demographics and ECG features as predictive variables with an AUROC of 0.843. For a fair comparison, we report an AUROC of 0.906 as a broad category heart failure, however, stress that our model provides very accurate differential diagnoses for acute/chronic systolic/diastolic heart failure. \cite{Attia2019b} report an AUROC score of 0.932 for the detection of a low ejection fraction of less than 35\%, which is often associated with heart failure. To put this into perspective, for heart failure with reduced ejection fraction (I50.20) we report an AUROC of 0.936. \citeSM{vidt2005hypertensive} review ECG changes related to \textit{hypertensive heart disease}. We report 0.814 AUROC for HCKD detection, and 0.929 for a combination of hypertensive heart and chronic kidney disease. \citeSM{schoenenberger2008prediction} predicted hypertensive crisis (HC) from 24-hour ambulatory blood pressure monitoring, while \citeSM{HOFFMAN2021100250} predicted hospital readmissions for HC specifically for pregnancy with AUROC of 0.85. We report 0.881 for HC detection. \citeSM{liu2022artificial} predicted pulmonary hypertension from ECGs and transthoracic echocardiography with an AUROC of 0.88, while \citeSM{ARAS20231017} reported 0.89, we report 0.890 for its detection. \citeSM{zhao2020early} differentiated ST-elevation \textit{myocardial infarction} (STEMI) from controls based on ECG data with 0.995 AUROC, whereas \citeSM{gustafsson2022development} achieved 0.991. We report a global performance of 0.870, however, we achieved higher performance on more fine-grained settings such as STEMI of the anterior wall and inferior wall of 0.986 and 0.950 respectively. Similarly, \citeSM{gustafsson2022development} achieved 0.832 for non-ST elevation (NSTEMI) whereas we achieved 0.847.

\xheading{Predictability of non-cardiac conditions}
We focus on non-cardiac conditions from selected chapters. A more detailed discussion across all chapters is in the supplementary material. Direct comparisons are limited due to scarce literature results. One such example is \citeSM{kwon2021deep}, who report an AUROC score of 0.901 for detecting \textit{sepsis}. Here we achieve an AUROC score of 0.86 across different kinds of sepsis in a more comprehensive patient collective. \cite{Kwon2020} demonstrated the feasibility of predicting \textit{anemia} from the ECG, reporting an AUROC score 0.923. We report AUROCs up to 0.857 for different sorts of anemias based on diverse causes. \cite{ahn2022development} have developed predictive models for the detection of cirrhosis from ECGs with an AUROC of 0.908, we report 0.906 AUROC for cirrhosis detection as well as 0.973 for cirrhosis with ascites. Also \textit{chronic kidney disease} has been shown to lead to ECG changes in patients with hypoalbuminemia \citeSM{10.3389/fcvm.2022.895201} and has been previously detected from ECGs with an AUROC of 0.767 \citeSM{holmstrom2023deep}, we report an AUROC of 0.834. Another work predicted mitral valve prolapse from ECGs either congenital or not with an AUROC of 0.80 \citeSM{PMID:37936601}. We report an AUROC of 0.863 for \textit{congenital malformations of aortic and mitral valves}. 

\subsection{Literature comparison: Cardiovascular conditions}

We proceed by discussing ICD codes at the 3-digit level that can be predicted accurately: 

\xheading{I25: Chronic ischemic heart diseases}
In the domain of \textit{chronic ischemic heart diseases}, \citeSM{ambale2017cardiovascular} used imaging noninvasive tests, and biomarker panels for the detection of atherosclerotic heart disease. \citeSM{neagoe2003neuro} uses a neuro-fuzzy binary classification setting for ischemic heart disease detection using ECG signals, specifically features extracted from the QRST zone. Remarkably, the study achieves a high recognition score, however, they do not use a performance metric that allows for a direct comparison to our results. 
We report AUROCs of 0.827 for chronic ischemic heart disease as a broad category, as well as 0.825, and 0.903 for atherosclerotic heart disease with native coronary artery without angina pectoris and with angina pectoris, respectively.

\xheading{I48: Atrial fibrillation}
\citeSM{Sehrawat2022} review diverse deep-learning methods for \textit{atrial fibrillation} and flutter detection by ECGs, which includes the work of \citeSM{kashou2020comprehensive}, with AUROCs ranging from 0.82 to 0.99. Deep learning models \cite{Strodthoff:2020Deep} on open source datasets such as PTB-XL \cite{Wagner:2020PTBXL} reach AUROC scores of 0.982. This should be compared to a score of 0.966 for persistent atrial fibrillation obtained from our model. Furthermore, \citeSM{10.1093/ehjdh/ztac014} used a deep learning model to predict atrial fibrillation from 24-h Holter recording, as well as \citeSM{10.1093/ehjdh/ztad018} based on RR-intervals, however, neither of them reports AUROC scores. In a more fine-grained setting, \citeSM{amaya2022development} reported an AUROC of 0.876 for \textit{paroxysmal atrial fibrillation}, in comparison to 0.891 from our model. Further, we report 0.966 for persistent AF, and 0.948 for chronic AF.

\xheading{I50: Heart failure}
\citeSM{hussain2020detecting} used temporal, spectral, and complex HRV dynamics to detect \textit{heart failure} with an AUROC of 0.97. \citeSM{doi:10.4070/kcj.2018.0446} used demographics and ECG features as predictive variables and reached an AUROC of 0.843. We report an AUROC of 0.906 as a broad category heart failure, however, stress that our model provides very accurate differential diagnoses for acute/chronic and systolic/diastolic heart failure. \cite{Attia2019b} report an AUROC score of 0.932 for the detection of a low ejection fraction <35\%, which is often associated with heart failure. For heart failure with reduced ejection fraction (I50.20) we report an AUROC of 0.936. 

\xheading{I11-I16: Hypertensive heart diseases}
\citeSM{vidt2005hypertensive} review ECG changes related to \textit{hypertensive heart disease}. 
\citeSM{martins2012hypertensive} investigated strategies to improve care for patients with hypertensive chronic kidney disease of certain (HCKD) ethnicity. We report 0.814 AUROC for HCKD detection, and 0.929 for a combination of hypertensive heart and chronic kidney disease. \citeSM{schoenenberger2008prediction} predicted hypertensive crisis (HC) from 24-hour ambulatory blood pressure monitoring, while \citeSM{HOFFMAN2021100250} predicted hospital readmissions for HC specifically within pregnancy with AUROC of 0.85. We report 0.881 for HC detection. \citeSM{liu2022artificial} predicted pulmonary hypertension from ECGs and transthoracic echocardiography with an AUROC of 0.88, while \citeSM{ARAS20231017} reported 0.89, we report 0.890 for its detection.

\xheading{I21 Acute myocardial infarction}
\citeSM{zhao2020early} differentiated ST-elevation \textit{myocardial infarction} (STEMI) from controls based on ECG data with 0.995 AUROC, whereas \citeSM{gustafsson2022development} achieved 0.991. We report a global performance of 0.870, however, we achieved higher performance on more fine-grained settings such as STEMI of the anterior wall and inferior wall of 0.986 and 0.950 respectively. Similarly, \citeSM{gustafsson2022development} achieved 0.832 for non-ST elevation (NSTEMI) whereas we achieved 0.847. Finally, myocardial infarction type 2 represents a more challenging diagnosis to differentiate and also increases the risk of post-discharge mortality \citeSM{10.1177/2048872614538411}, for which we achieved 0.936. Further, \citeSM{strodthoff2018detecting} also investigated myocardial infarction however with different evaluation metrics.

\xheading{I07,I08,I34,I35: Valve disorders}
With respect to \textit{valve disorders}, \citeSM{PMID:36788316} achieved 0.88 for mitral valve prolapse, whereas \citeSM{tison2019automated} reported 0.77. The proposed model reaches an AUROC score of 0.913. \citeSM{PMID:36788316} achieved 0.89 on aortic stenosis, where \citeSM{Haimovich2023} correlates LVH with aortic stenosis with an AUROC of 0.92. We achieve 0.879. Further, we achieve 0.948 for tricuspid valve diseases, and significant performance on disorders on multiple valves such as mitral and artic with 0.863, as well as mitral and tricuspid with 0.944. 

\xheading{I44,I45: Conduction disorders}
The predictive power of the ECG for the prediction of \textit{atrioventricular blocks} (AV) has been studied prominently in the literature \cite{Ribeiro2020}, however, they only report global model performance which includes more than these conditions. \cite{Strodthoff:2020Deep} report 0.971, 0.860, and 0.995 for 1st, 2nd, and 3rd degree whereas we report 0.908, 0.953, and 0.957 respectively. Similarly, other blocks such as left and right bundle-branch (LBBB) and (RBBB) have previously reported significant predictive performance in the literature, where \citeSM{sadeghi2022deep} achieved 0.875 and \cite{Strodthoff:2020Deep} 0.998 for LBBB while we report 0.976. \citeSM{jeong2021convolutional} achieved 0.93 and \cite{Strodthoff:2020Deep} 0.998 for RBBB while we achieved 0.964.

\xheading{I65-I69: Cerebrovascular issues}
Imaging modalities are commonly used to characterize the anatomy of \textit{cerebrovascular issues} such as carotid issues and justify open surgical interventions, see \citeSM{HECK202149} for a review. From the ECG alone, we can detect carotid artery occlusion and stenosis with 0.836, cerebral atherosclerosis with 0.800, and sequels of cerebral infarction with 0.877, with fine-grained diagnosis after the infarction such as hemiplegia and hemiparesis with 0.873.

\subsection{Literature comparison: Non-cardiovascular conditions}
We structure the discussion of non-cardiovascular conditions by ICD-10 chapters: 

\xheading{I: Certain infectious and parasitic diseases} \citeSM{kwon2021deep} presented an algorithm to detect \textit{sepsis} with a high predictive accuracy (AUROC 0.901). Here, we achieve an AUROC score of 0.86 across different kinds of sepsis in a more realistic patient collective. There are no literature results on the direct prediction of specific infectious diseases from the raw ECG (e.g. \textit{E.coli} with AUROC 0.862).

\xheading{II: Neoplasms} Also the prediction of \textit{neoplasms} has not been investigated directly from ECG data, even though isolated records on ECG changes due to the presence of lung cancer exist \citeSM{ASTORRI2000225}. Similarly, \citeSM{spinu2021ecg} investigated the sub-clinical heart damage and ECG abnormalities due to the toxic chemotherapy effect, as well as \citeSM{pohl2020ecg} for melanoma patients, which might in fact also represent a confounding factor in our case. We demonstrate significant predictive performance on neoplasms, primarily in the bronchus and lungs 0.827, brain 0.827, prostate 0.802, pancreas 0.884, and bone marrow 0.933, as well as \textit{leukemia} with 0.922.

\xheading{III: Diseases of the blood and blood-forming organs and certain disorders involving the immune mechanism} \cite{Kwon2020} demonstrated the feasibility of predicting \textit{anemia} from the ECG, reporting an AUROC score 0.923, whereas we have AUROCs up to 0.857 for different sorts of anemias based on diverse causes. Nevertheless, we also report significant AUROCs for the disease of the blood and blood-forming organs such as 0.812 for \textit{pancytopenia} and 0.933 for \textit{disseminated intravascular coagulation}.

\xheading{IV: Endocrine, nutritional and metabolic diseases } \textit{Diabetes} represents the dominant condition in the endocrine, nutritional, and metabolic chapter, which is known to impact the ECG \citeSM{stern2009ecg,swapna2020diabetes}. \cite{kulkarni2023machine} developed an algorithm that was able to discriminate between no-diabetes, pre-diabetes and type 2 diabetes, but did not report AUROC scores. We report significant prediction accuracy both on diabetes type 1 and 2 with diverse secondary complications, e.g. type 2 with kidney complications 0.924 and neurological complications 0.880. Further, we also achieve high performance on sterol and serum electrolyte conditions such as \textit{cholesterolemia}, \textit{calcemias}, \textit{kalemias} and \textit{magnesemias}.

\xheading{V: Mental and behavioural disorders } We report an AUROC of 0.819 for the direct detection of dementia from raw ECG signals, whereas prior work already demonstrated the feasibility of inferring dementia risk within a horizon of 5 years based on the ECG \citeSM{ISAKSEN2022106640}. Similarly, we also report high predictive performance for diverse \textit{substance disorders and dependencies} such as opioids, cocaine, alcohol, and tobacco, which are know for leaving traces in the ECG \citeSM{SERENY1971}.

\xheading{VI: Diseases of the nervous system} Recent studies \citeSM{karabayir2023externally} highlight the potential to predict \textit{Parkinson's disease} (PD) from ECGs achieving an AUROC score of 0.74 up to 1 year before diagnosis. We report 0.803 AUROC for PD (after diagnosis). \textit{Alzheimer's}: Our AUROC score of 0.803 (after detection) is in line with this finding. Previous studies demonstrate certain ECG abnormalities on patients with \textit{polyneuropathy} \citeSM{beckman1992electrocardiographic}, we report 0.832 AUROC for its detection. 
Finally, we also found a certain degree of significance for \textit{anoxic brain damage} with 0.813 AUROC, which to the best of our knowledge has not been investigated before.

\xheading{X: Diseases of the respiratory system } \textit{Respiratory failure} correlations with ECGs have previously been investigated based on abnormal ECG features in pediatrics prognosis \citeSM{keith1961electrocardiogram}, the presence of cardiac arrhythmias in respiratory failure from chronic obstructive pulmonary disease \citeSM{incalzi1990cardiac} and respiratory signal extraction from ECGs \citeSM{travaglini1998respiratory}. We report AUROCs of 0.923 for acute and chronic respiratory failure and 0.951 with the additional presence of \textit{hypoxia}. Specific ECG feature abnormalities were found in a single patient with \textit{malignant pleural effusion} \citeSM{manthous1993pleural}, we report an AUROC of 0.904 for malignant pleural effusion detection. A similar study displayed specific heart conditions and ECG signal abnormalities with the presence of \textit{bronchiectasis} \citeSM{alhamed2021ecg}, we report an AUROC of 0.803, as well as 0.817 for \textit{atelectasis}, 0.883 for \textit{interstitial pulmonary diseases}, and 0.801 for \textit{asthma}.

\xheading{XI: Diseases of the digestive system } Recent studies around liver conditions such as \textit{cirrhosis} have been able to differentiate ECG changes between patients with cirrhosis and hepatitis \citeSM{toma2020electrocardiographic}, as well developing predictive models for its detection \cite{ahn2022development} with AUROC of 0.908. We report 0.906 AUROC for cirrhosis detection as well as 0.973 for cirrhosis with ascites. Furthermore, we also are able to predict with significant AUROCs diverse additional conditions which to the best of our knowledge, have not been investigated in similar predictive settings such as \textit{alcoholic hepatitis} 0.930, \textit{hemorrhagic necrosis of liver} 0.948, \textit{hepatic failure} 0.947, \textit{gastroparesis} 0.867, \textit{peritonitis} 0.853, and \textit{cholangitis} 0.848.

\xheading{XII: Diseases of the skin and subcutaneous tissue } The detection of \textit{ulcers} by ECG has not been investigated in detail, however, there are some works that found a correlation between ECG changes and upper gastrointestinal bleeding \citeSM{fa2006electrocardiographic}, as well as peptic ulcer detection from ECG and additional predictors such as respiration rate, heart rate, pH of saliva, and temperature \citeSM{faiz2023aiding}. We report significant AUROCs on \textit{pressure ulcers} at different locations such as back 0.975, and sacral region 0.874, as well as \textit{non-pressure ulcers} such as heel and midfoot 0.87, lower limb 0.819, and foot 0.811. Similarly, for \textit{acute lymphangitis} we report an AUROC score of 0.851.

\xheading{XIII: Diseases of the musculoskeletal system and connective tissue } \textit{Gout}, specifically serum uric acid levels, have been previously linked to ECG abnormalities \citeSM{cicero2015relationship}. 
At the time of the publication, we are not aware of any work that detects gout by a single ECG, a task for which we report an AUROC score of 0.803. Similarly, \textit{lupus} have been found to affect ECG signals \citeSM{hu2022prevalence}. This is in line with an AUROC score of 0.801 for lupus detection from ECG achieved by our model. Our robust approach also allows us to detect diverse musculoskeletal and tissue diseases such as \textit{arthropath} with 0.847, \textit{cartilage disorder} with 0.819, and \textit{disorders of bone density and structure} with 0.804 of AUROCs.

\xheading{XIV: Diseases of the genitourinary system } \textit{Chronic kidney disease} has been previously detected through ECGs with an AUROC of 0.767 \citeSM{holmstrom2023deep}, we report an AUROC of 0.834. \textit{Benign prostatic hyperplasia} have been shown to increase the incident of atrial fibrillation \citeSM{koccak2022relationship}, we report an AUROC of 0.811 for its detection. Patients with reduced glomerular filtration rate are more likely to have a history of cardiovascular diseases \citeSM{10.1001/archinte.164.9.969}, we report an AUROC of 0.878 for the detection of \textit{glomerular disease}. Kidney failure, particularly in end-stage renal disease patients is associated with an increased risk of sudden cardiac death by low heart rate variability \citeSM{10.1093/ndt/gfm634}, we report an AUROC of 0.852 for \textit{kidney failure with tubular necrosis} detection. 

\xheading{XVII: Congenital malformations, deformations and chromosomal abnormalities} Although congenital heart valve malformations are one of the most common types of birth defects, their severity is diverse. Most of these malformations are diagnosed by echocardiograms, however, early works provided a review that summarizes specific ECG abnormalities for various malformations \citeSM{khairy2007clinical}. Similarly, another work predicted mitral valve prolapse from ECGs either congenital or not with an AUROC of 0.80 \citeSM{PMID:37936601}. We report an AUROC of 0.863 for  \textit{congenital malformations of aortic and mitral valves}.

\xheading{XVIII: Symptoms, signs and abnormal clinical and laboratory findings} It has been shown that systemic inflammation increases the risk for atrial fibrillation \citeSM{lazzerini2019systemic}, we report an AUROC of 0.870 for \textit{systemic inflammation} detection. Although \textit{cardiogenic shock} is difficult to diagnose due to variable presentations, overlapping with other shock states, and specialized test requirements, previous works were able to identify it at ED by ECGs and ultrasounds \citeSM{DALY20202425}.
We report a 0.931 AUROC for its detection. To the best of our knowledge, \textit{cachexia} has not been detected by an ECG alone, however, as it is linked to an increased risk of arrhythmias, and can cause electrolyte imbalances it seems a possible task to achieve, for which we report an AUROC of 0.900.

\xheading{XIX: Injury, poisoning and certain other consequences of external causes } \textit{Patient poisoning} due to administrated drugs can cause diverse effects ranging from rash to brain damage, coma, or death. A previous work \citeSM{yates2012utility} explored with a controlled study the changes that ECGs suffer from patients with acute cardiotoxicity, classifying these by specific ion channels.
Based on the nature of our ICD-10 labeling, we report significant AUROCs for diverse poisoning based on groups of drugs, such as anticoagulants  0.880, antibiotics 0.872, antineoplastic and immunosuppressive drugs 0.870, systemic and hematological agents 0.851, hormones 0.841, diuretics 0.824, and opioids 0.817. Similarly, as injury we report 0.8345 of AUROC for \textit{pertrochanteric fracture}.

\xheading{XX: External causes of morbidity and mortality } In the domain of \textit{traffic accidents}, a prior study \citeSM{SHAIKHALARAB2012790} measured survivors' HRV within 24 hours post-accident, at 2 months, and 6 months thereafter. The findings underscore a consistent diminution in HRV over time. Similarly, other works investigated specific body damaged areas such as blunt cardiac injuries \citeSM{kyriazidis2023accuracy}, however, concluded that an ECG on its own does not achieve significant sensitivity. Surprisingly, our model excels in predicting various ICD-10 codes associated with \textit{road traffic and non-traffic accidents} with high AUROCs.

\xheading{XXI: Factors influencing health status and contact with health services } With regard to health status, a previous work explored ECG changes across different \textit{body mass indexes} of healthy individuals \citeSM{hassing2019body}.

Our work shows high discriminative power for diverse indexes, especially for extreme ones such as <19.9 and 50-59 with 0.944, and 0.940 AUROCs, respectively. Surprisingly, our model predicts with 0.841 AUROC the \textit{absence of limb}, which highlights possible research directions in terms of limb prosthesis acceptance or rejection, for which a previous work based on patient metadata found significant 
results \citeSM{biddiss2008multivariate}. Similarly, our models report an AUROC of 0.935 for the \textit{encounter for adjustment and management of implanted cardiac devices}, which also highlight possible research directions and application in terms of the reduction of adverse events with patient timely intervention by remote device monitoring.

\begin{figure*}[ht]
    \centering
    \includegraphics[width=0.99\textwidth]{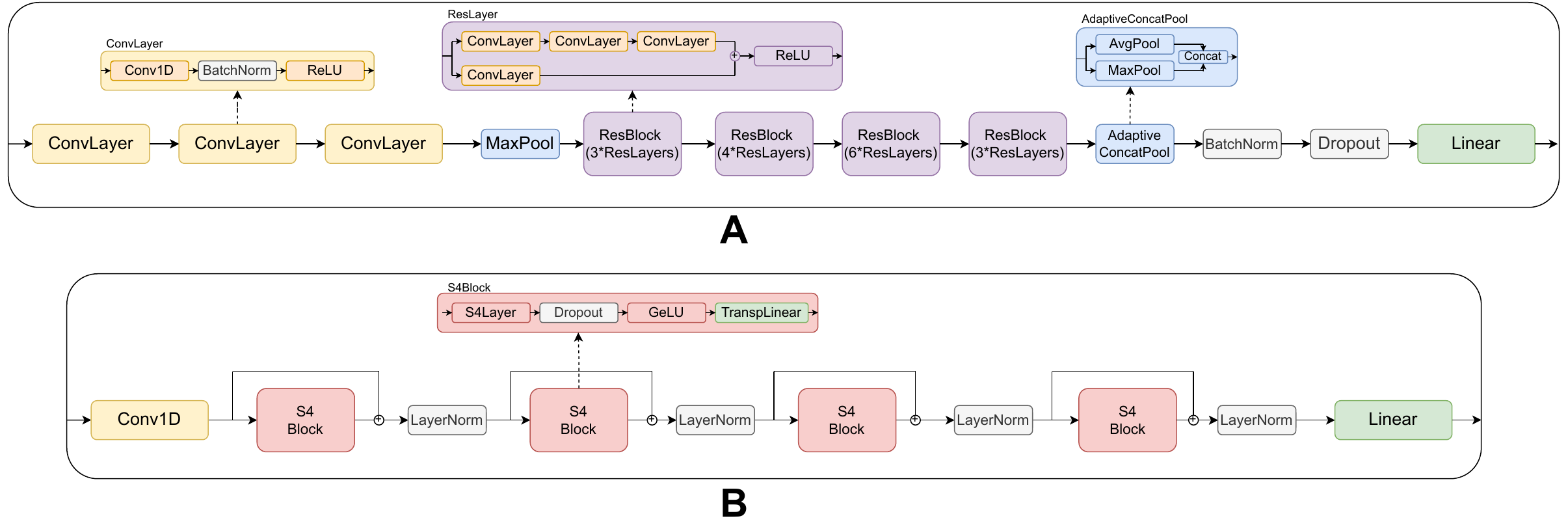}
    \caption{Schematic representation of the two model architectures considered in this work: (A) XResNet1d50 (B) S4-model.}
    \label{figure:architecture}
\end{figure*}

\subsection{Model architectures} In this work, we consider two model architectures, an example of a state-of-the-art convolutional model as a representative of this most widely model architecture and a structured state space model as a contender that showed statistically significant improvements over convolutional architectures in a previous study. We summarize both model architectures graphically in \Cref{figure:architecture}.

\heading{XResNet1d} A recent benchmarking study on the PTB-XL dataset \cite{Strodthoff:2020Deep} established the superiority of modern ResNet- or Inception-based convolutional architectures over other deep learning-based architectures and feature-based approaches. This is in line with the use of ResNet as the predominant model architecture for deep-learning-based ECG analysis. The XResNet1d family introduced in \cite{Strodthoff:2020Deep} is a family of convolutional models as adaptations of the popular XResNet models \citeSM{he2019bag} from image recognition to one-dimensional data. In particular, the XResNet1d50 architecture considered here is a convolutional neural network for 1D data that consists of 4 residual blocks, each with a set of 3,4,6, and 3 convolutional layers. 

\heading{S4} Structured state space models represent a recently proposed alternative \cite{gu2021efficiently} to convolutional architectures which demonstrated compelling abilities in capturing long-range dependencies in sequential data such as time series, including physiological time series. It was used as a building block for time series imputation and forecasting \citeSM{Alcaraz:2022Diffusion} also covering applications to ECG data as well as the generation of synthetic ECG data \citeSM{ALCARAZ2023107115}. A recent benchmarking study \cite{Mehari2023S4} established statistically significant improvements in this architecture over the existing state-of-the-art of mostly convolutional and recurrent architectures also for diagnostic ECG tasks, both in the supervised and in the self-supervised setting. Here, we closely follow the hyperparameter choices from \cite{Mehari2023S4}, who used four bidirectional S4-layers with a model dimension of 512 and a state dimension of 8. The S4-layers preserve the temporal resolution of the input and can be seen as a replacement for a transformer, an RNN layer, or a convolutional layer with unit stride. The S4 layers are followed by an average pooling layer and a linear classification head.

\subsection{MIMIC-IV-ECG-ICD-ED dataset}

\begin{figure*}[ht]
    \centering
    \includegraphics[width=0.50\textwidth]{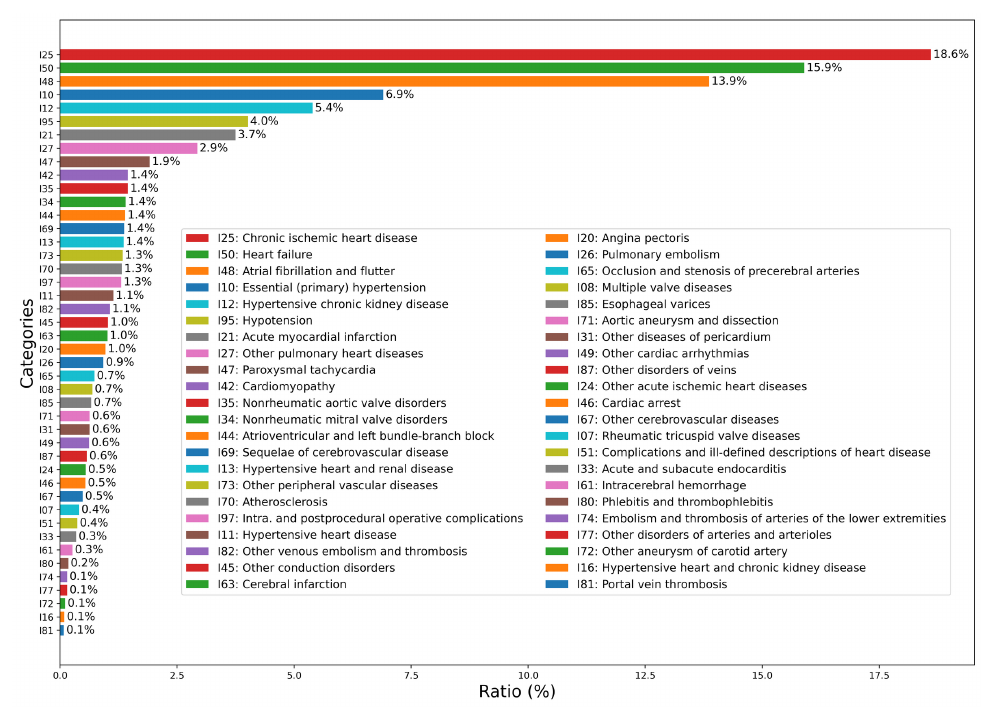}
    \caption{Representation of the distribution of cardiac conditions in the ED dataset within chapter IX (Circulatory system diseases categories) at the 3rd digit level including all of its descendants, where category I25 (Chronic ischemic heart disease) is the most represented category with 19.6\%, closely followed by I50 (heart failure) with 15.3\%, I48 (Atrial fibrillation and flutter) with 13.5\%, and I10 (Essential hypertension) with 10.7\%.}
    \label{figure:distribution_cardiac_ed}
\end{figure*}

\begin{figure*}[ht]
    \centering
    \includegraphics[width=0.99\textwidth]{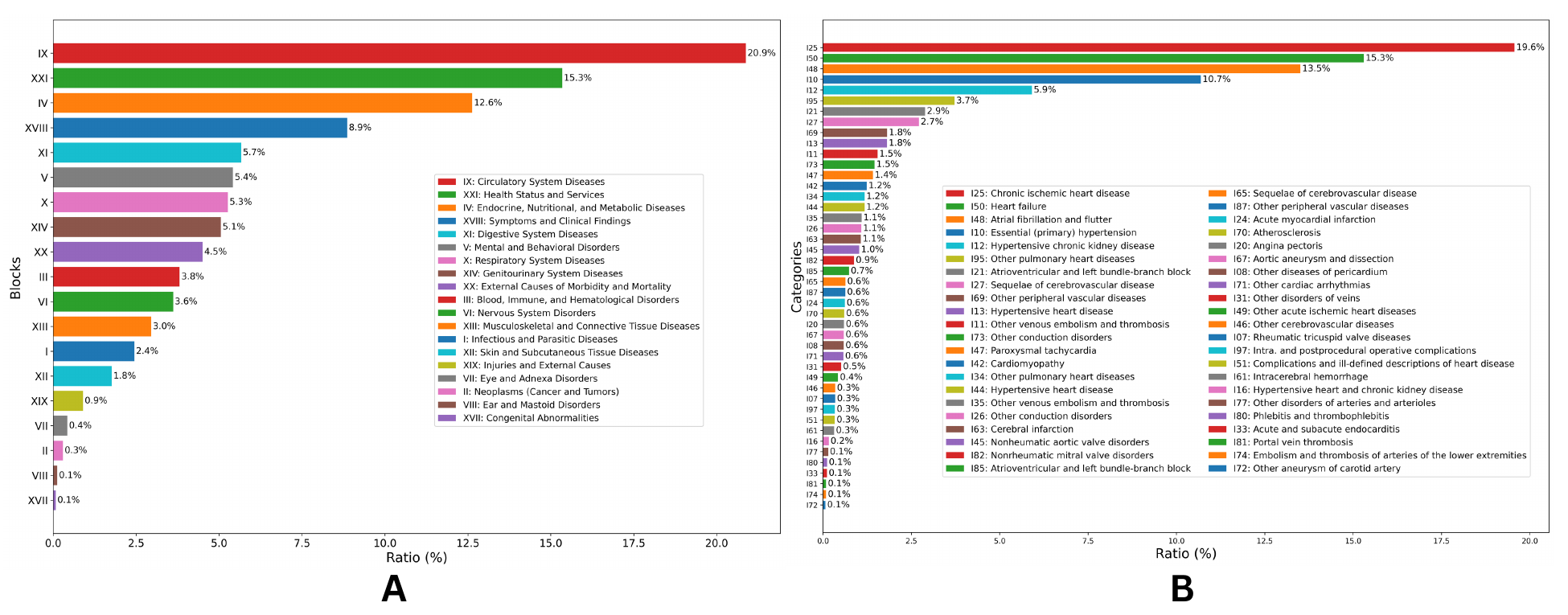}
    \caption{Representation of the distribution of the full dataset instead of just the ED subset: Descriptive statistics on the MIMIC-IV-ECG-ICD dataset obtained by joining the MIMIC-IV-ECG with diagnoses from the clinical MIMIC-IV dataset: (A) represents the distribution of chapters, whereas (B) represents the distribution with the cardiovascular chapter IX.}
    \label{figure:descriptive_all}
\end{figure*}

In this section, we describe the ICD10 chapters and distributions of the processed MIMIC-IV-ECG database \cite{Johnson2023}\citeSM{MIMICIV}. In \Cref{figure:descriptive_all}, we summarize the full dataset in terms of the distribution of chapters and IX categories. \Cref{figure:descriptive_all}A represents the distribution of chapter records, where chapter IX is most strongly represented chapter with 20.9\%, closely followed by chapter XXI with 15.3\%, chapter IV with 12.6\%, and chapter XVIII with 8.9\%. \Cref{figure:descriptive_all}B shows the distribution of  codes within the cardiovascular chapter IX aggregated at the 3rd digit level with all of its descendants included, where category I25 is the most represented category with 18.6\%, closely followed by I50 with 15.9\%, I48 with 13.9\%, and I10 with 6.9\%. 

\subsection{Evaluation scenarios}

\xheading{Disagreement between ED and hospital discharge diagnoses} Previous studies evaluated the agreement between diagnoses assigned in ED and those assigned after hospital discharge \citeSM{Farchi2007}. Some of their findings demonstrate that only 62.2\% of ED diagnoses were concordant with hospital discharge diagnoses primarily for older patients and less urgent cases. However, there were other concordance factors such as an hour of the visit, and ED specialization degree. Thus, the correct diagnosis assignment at the ED represents a challenge not only as a system overall but also facility-wise, and it should be overcome as it has been shown that diagnoses that changed from ED to hospital discharge (wrong diagnosis at the ED department) had a 30\% higher probability of death \citeSM{Farchi2007}. For this work, we accept a potential mismatch between ED diagnoses and hospital discharge diagnoses and take it into consideration during the model building process.

\xheading{Comprehensive model T(ALL2ALL)} We first want to introduce a training-testing scenario which we refer to as T(ALL2ALL)-E(ALL2ALL). The first half of the acronym refers to the training modality: The model was trained on all ECGs (hospital+ED), to predict all hospital discharge diagnoses (if available) otherwise ED diagnosis. The second half represents the evaluation scenario: The model was evaluated on all ECGs (hospital+ED) to predict all hospital discharge diagnoses (if available) otherwise ED diagnoses. In this scenario, both training and test distribution coincide both in terms of patient population as well as label selection. From the training T(ALL2ALL) strategy, we also consider four additional evaluation scenarios which differ based on the ECGs and department diagnosis subsets: T(ALL2ALL)-E(ALL2HOSP), T(ALL2ALL)-E(ED2ALL), T(ALL2ALL)-E(ED2HOSP), T(ALL2ALL)-E(ED2ED). The diagnoses label set in the evaluation scenario either for ED (2ED), or hospital (2HOSP) is restricted to these sets in contrast to all (2ALL).

\xheading{Application scenarios} The scenarios T(ALL2ALL)-E(ALL2ALL) and T(ALL2ALL)-E(ALL2HOSP) address the question of inferring general conditions from raw ECGs. Discharge diagnoses as compared to ED diagnoses represent the more comprehensive set of ICD codes with a potentially higher label certainty. However, the omission of ED ECGs without hospital admission in the T(ALL2ALL)-E(ALL2HOSP) scenario leads to a potentially severe bias in the target population as the subset of ED patients who did not get admitted to the hospital represents a large fraction of cases with typically less severe symptoms. The T(ALL2ALL)-E(ED2ALL), T(ALL2ALL)-E(ED2HOSP), and T(ALL2ALL)-(ED2ED) scenarios can be seen as potential applications for a triage use case. One can either predict the most comprehensive label set available, i.e., hospital discharge diagnosis if available or otherwise ED diagnosis or use only hospital diagnosis or ED diagnosis as targets. Note that the choice of hospital diagnoses (in scenarios T(ALL2ALL)-E(ALL2HOSP) and T(ALL2ALL)-E(ED2HOSP)) leads to a substantial reduction in terms of applicable ED ECGs as all ECGs without subsequent hospital admission get excluded.

\xheading{Distribution shift} On the one hand, using an evaluation scenario that deviates from the training scenario T(ALL2ALL) inevitably leads to a mismatch between training and test distributions. On the other hand, one might expect that the models on the smaller selections profit from the training on the largest available dataset in scenario, i.e.,  T(ALL2ALL). However, our experiments demonstrate in terms of summary metrics such as macro AUROC that the advantage of training on a training dataset that matches the test distribution outperforms a model trained on potentially significantly larger training dataset. Therefore we present a model T(ED2ALL)-E(ED2ALL) that during training and evaluation only uses training ED ECGs to predict discharge diagnosis (if available) otherwise ED diagnosis, with an additional evaluation scenario T(ED2ALL)-E(ALL2ALL) which evaluated on all ECG subsets to predict discharge diagnosis (if available) otherwise ED diagnosis. Furthermore, we also present a model T(ED2ED)-E(ALL2ALL) trained on ED ECGs to predict ED labels and is evaluated on both subsets of ECGs to predict discharge diagnosis (if available) otherwise ED diagnosis.

\begin{table}[!ht]
    \centering
    \begin{tabular}{ll}
    \toprule
    Scenario     &  macro AUROC \\
    \midrule
    \midrule
    \textbf{T(ALL2ALL)-E(ALL2ALL)} & \textbf{0.7505 (0.7478-0.7531)}  \\
    T(ED2ALL)-E(ALL2ALL) & 0.7335 (0.7309-0.7366)  \\ 
    T(ED2ED)-E(ALL2ALL) & 0.6777 (0.6736-0.6816)   \\ 
    \midrule
    T(ALL2ALL)-E(ED2ALL) &  0.7691 (0.7651-0.7732)  \\
    \textbf{T(ED2ALL)-E(ED2ALL)} & \textbf{0.7742 (0.7703-0.7783)}   \\ 
    \midrule   
    T(ALL2ALL)-E(ALL2HOSP) & 0.7301 (0.7273-0.7330) \\ 
    \midrule
    T(ALL2ALL)-E(ED2HOSP) & 0.7394 (0.7350-0.7433) \\
    \midrule
    T(ALL2ALL)-E(ED2ED) & 0.6589 (0.6416-0.6568)  \\
    \bottomrule
    \end{tabular}
    \caption{Model performance (S4-model) in terms of macro AUROC scores across different training and evaluation scenarios. The models highlighted in boldface correspond to the best-performing model for a given evaluation scenario. In the main text, the setup (ED2ALL)-E(ED2ALL) was used. Note that only scenarios with identical evaluation setups (separated by horizontal lines) are directly comparable.}
    \label{tab:all_scenarios}
\end{table}

\begin{figure*}[ht]
    \centering
    \includegraphics[width=1\textwidth]{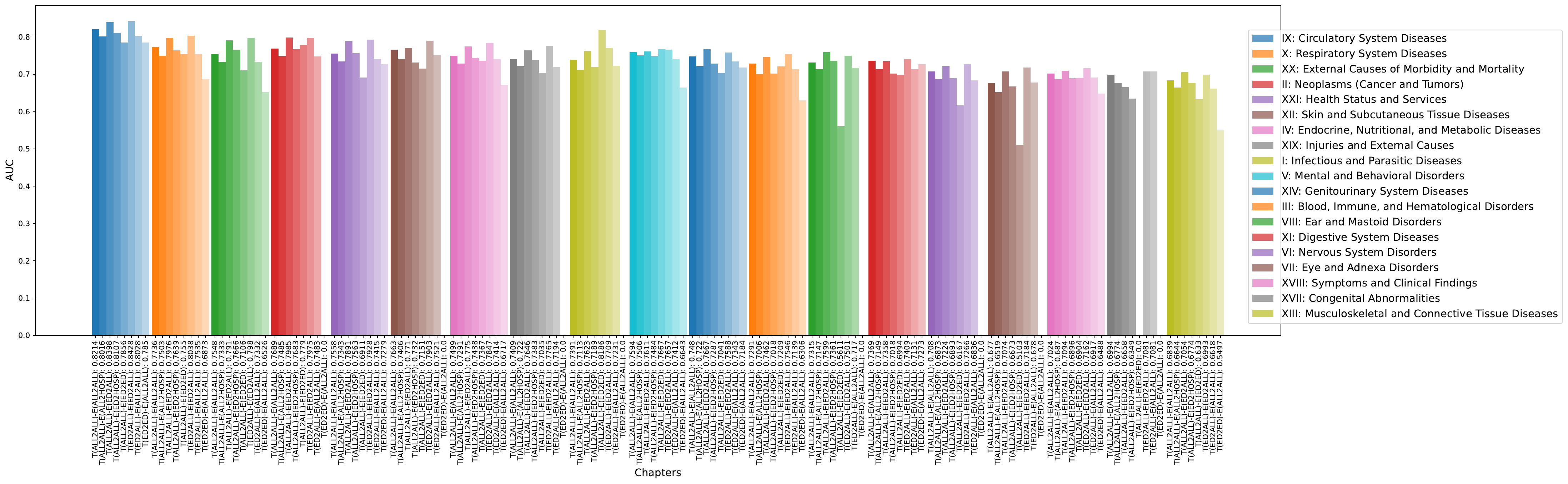}
    \caption{The bar plot dissects all our investigated scenarios, representing diverse training and testing subsets characterized by varying ECG sources (ED, Hosp, or ALL) as well as distinct label sets (ED, Hosp, or ALL). Each color group showcases the predictive performance across each of the ICD-10 chapters This nuanced representation provides a concise yet comprehensive snapshot of health patterns across different scenarios, offering valuable insights into the impact of data source and label variation on predictive models for various medical conditions. The values chapters bar blocks are sorted by the T(ED2ALL)-E(ED2ALL) scenario. Similarly, the chapter's bar blocks are sorted left to right as they are presented in the legend box up to down.}
    \label{fig:all_scenarios}
\end{figure*}

\xheading{Performance comparison E(ED2ALL)} \Cref{fig:all_scenarios} shows the model performance in each of the scenarios along with macro AUROC and confidence intervals. The model training on the full dataset and evaluated  only  on ED ECGs, i.e.,  T(ALL2ALL)-E(ED2ALL), yields a model that is only slightly weaker than a specialized model only trained on ED data, i.e., T(ED2ALL)-E(ED2ALL) (0.769 vs. 0.774). Nevertheless, as their scores overlap within error bars, we assessed the statistical significance of the performance difference using bootstrapping on the test set, where a confidence interval for the score difference that does not overlap with zero indicates a statistically significant result. The global model T(ALL2ALL)-E(ED2ALL) reported 76 conditions as significant while the T(ED2ALL)-E(ED2ALL) model 105, both models achieved 0.7691 and 0.7742 of macro AUROC respectively, where the difference also turns out significant, therefore the specialized model T(ED2ALL)-E(ED2ALL) is stronger. However, it is also worth stressing that the specialized ED model is considerably weaker on the set of all ECGs T(ED2ALL)-E(ALL2ALL) vs. T(ALL2ALL)-E(ALL2ALL) (0.734 vs. 0.751). \Cref{fig:all_scenarios} shows the macro AUROC for each of the scenarios summarized according to ICD-10 chapters, providing a more comprehensive comparison of specific sorts of conditions for each of the possible clinical applications among departments. Bars with macro AUROC of 0 do not contain conditions at the specific chapter due to the absence of these during the label selection stage.  Overall, the T(ED2ALL)-E(ED2ALL) model is best on all chapters except for II (Neoplasms), XIV (Diseases of the genitourinary system), VIII (Diseases of the ear and mastoid process), and XIII (Diseases of the musculoskeletal system and connective tissue). Furthermore, it is important to note that the additional scenario T(ALL2ALL)-E(ED2ED) is better than the two of them on chapters I (Certain infectious and parasitic diseases) and V (Mental and behavioural disorders).

\xheading{Performance comparison E(ALL2ALL)} The second scenario we investigate in more detail is the evaluation on both hospital and ED ECGs. To this end, we first compare T(ALL2ALL)-E(ALL2ALL), T(ED2ALL)-E(ALL2ALL), T(ED2ED)-E(ALL2ALL), which all evaluate all diagnoses from all ECGs (global perspective of any ECG), however, while trained on all records with all labels (global perspective), ED records with all labels (ED+Hospital screening use case at the ED department), and ED records on ED labels (ED screening use case at the ED department). Overall, the comprehensive model T(ALL2ALL)-E(ALL2ALL) is best on all chapters except for VII ( Diseases of the eye and adnexa), and XIII (Diseases of the musculoskeletal system and connective tissue), where T(ED2ALL)-E(ALL2ALL) is better. Similarly, T(ED2ALL)-E(ALL2ALL) outperforms T(ED2ED)-E(ALL2ALL) in all ICD chapters except for XI (Diseases of the digestive system).

\begin{table}[!ht]
    \centering
    \begin{tabular}{cccc}
    \toprule
    Statement & ICD10 codes & Internal & External \\
    \midrule
    1AVB & I440 & 0.9007 & 0.9041 \\
    AFIB & I4891 & 0.8965 & 0.9872 \\
    LBBB & I447 & 0.9588 & 0.9990 \\
    RBBB & I4510 & 0.9384 &  0.9959 \\
    SBRAD & R001 & 0.7620 &  0.9368 \\
    STACH & R000 & 0.8305  &  0.9740 \\
    \bottomrule
    \end{tabular}
    \caption{External validation for diverse cardiac conditions on the E(ALL2ALL) setting (1AVB: 1st degree AV block, AFIB: atrial fibriallation, LBBB: left bundle branch block, RBBB: right bundle branch block, SBRAD: sinus bradycardia, STACH: sinus tachycardia).}
    \label{tab:external_1111}
\end{table}

\xheading{External validation E(ALL2ALL)}: \Cref{tab:external_1111} presents the external validation scores for diverse cardiac conditions in the E(ALL2ALL) setting, once again validating an optimal performance outside of the training dataset distribution.

\subsection{Comparing model architectures}

\begin{table}[ht]
    \centering
    \scalebox{1}{
    \begin{tabular}{llll}
    \toprule
    Model & Scenario & macro AUROC &  Significant labels  \\
    \midrule
     XResNet1d50 &T(ALL2ALL)-E(ALL2ALL) & 0.7475(0.7447-0.7504)&  43/1076     \\ 
     \textbf{S4} & T(ALL2ALL)-E(ALL2ALL) &\textbf{0.7504(0.7478-0.7531)}$\ast$& \textbf{161/1076}  \\ 
     \midrule
     XResNet1d50 &T(ED2ALL)-E(ED2ALL) & 0.7724 (0.7682-0.7761)&   42/1076          \\ 
     \textbf{S4} & T(ED2ALL)-E(ED2ALL) &\textbf{0.7742 (0.7703-0.7783)} &\textbf{104/1076}   \\ 
    \bottomrule
    \end{tabular}}
    \caption{Comparison of predictive performance between the XResNet1d50 and S4 models on the proposed MIMIC-IV-ECG-ICD-ED dataset. The table shows the macro AUROCs, and the resulting performances of a bootstrap-powered hypothesis test in terms of significant labels and macro AUROC for each model at the settings T(ALL2ALL)-E(ALL2ALL), and T(ED2ALL)-E(ED2ALL) with the S4 model outperforming the XResNet1d50 model in both scenarios. Confidence intervals for AUROC scores are provided in parentheses. The fourth column indicates the number of statements for which the model outperforms its competitor (in the sense of bootstrapping confidence intervals for the score difference not overlapping with zero). Bold values indicate the best-performing model in both scenarios. In both cases the performance difference between both models in terms of macro AUROC is statistically significant.}
    \label{tab:benchmarking}
\end{table}

In this section, we provide a comparative assessment of two model architectures that have shown competitive performance on the PTB-XL dataset, see \cite{Strodthoff:2020Deep,Mehari2023S4}. More specifically, we compare an XResNet1d50 model, a convolutional neural network with ResNet architecture, to an S4 model leveraging the recently proposed structured state-space models \cite{gu2021efficiently}, see Methods for details. As in \cite{Strodthoff:2020Deep}, we report AUROCs for the respective statements in the label set. \Cref{tab:benchmarking} shows the predictive performance on the T(ALL2ALL)-E(ALL2ALL) setting for the XResNet1d50 model, showcasing a macro AUROC of 0.7475, while the S4 model achieved a slightly higher AUROC of 0.7504. However, as their confidence intervals overlap, we assessed the statistical significance of the performance difference using bootstrapping on the test set. The XResNet1d50 model reported 43 labels as significant whereas the S4 model 161, both models achieved a macro AUROC on the hypothesis test of 0.7475 and 0.7505 respectively. Similarly, on the T(ED2ALL)-E(ED2ALL) setting, the XResNet1d50 model, showcased a macro AUROC of 0.7724, while the S4 model achieved a slightly higher AUROC of 0.7742. Again, we performed a bootstrap-powered hypothesis test to properly distinguish the significance of their performances. The XResNet1d50 model reported 42 labels as significant whereas the S4 model 104, both models achieved a macro AUROC on the hypothesis test of 0.7724 and 0.7774 respectively. Therefore, the S4 model outperforms the ResNet in terms of predictive performance in both scenarios.

\subsection{Complete results tables}

\begin{table}[ht]
\centering
\scalebox{0.60}{
\begin{tabular}{lp{10cm}p{10cm}}
\toprule
Block: Block description. Block AUROC & Code: Code AUROC. Code description &  Code: Code AUROC. Code description \\  
\midrule
\midrule

\multirow{3}{*}{I: Infectious and Parasitic Diseases. 0.7709} & B9620: 0.862. E. coli & A40: 0.859. Streptococcal sepsis \\ 
 & A41: 0.857. Other sepsis     &  B370: 0.823. Candidal stomatitis   \\
 &  A047: 0.809. Enterocolitis due to Clostridium difficile   &     \\ 
 
\midrule
\multirow{3}{*}{II: Neoplasms (Cancer and Tumors). 0.7975} & \textbf{C7952: 0.933. Malignant neoplasm of bone marrow}  & \textbf{C925: 0.922. Acute myelomonocytic leukemia}  \\        
 &  C25: 0.884. Malignant neoplasm of pancreas  & D469: 0.869. Myelodysplastic syndrome    \\
 &  C8589: 0.853. Other specified types of non-Hodgkin lymphoma  &  C34: 0.826. Malignant neoplasm of bronchus and lung    \\

\midrule
\multirow{3}{*}{III: Blood, Immune, and Hematological Disorders. 0.7546}& \textbf{D65: 0.933. Disseminated intravascular coagulation} & D684: 0.878. Acquired coagulation factor deficiency \\
 & D631: 0.857. Anemia in chronic kidney disease &  D618: 0.812. Other specified aplastic anemias and other bone marrow failure syndromes   \\

\midrule
\multirow{4}{*}{IV: Endocrine, Nutritional, and Metabolic Diseases. 0.7847}& \textbf{E1129: 0.924. Type 2 diabetes mellitus with diabetic kidney complication}   &   \textbf{E660: 0.907. Obesity due to excess calories}  \\
 &  E103: 0.899. Type 1 diabetes mellitus with ophthalmic complications  &  E43: 0.886. Severe protein-calorie malnutrition   \\
 & E1342: 0.880. Diabetes mellitus with diabetic polyneuropathy  & E8770: 0.874. Fluid overload \\
 
\midrule
\multirow{3}{*}{V: Mental and Behavioral Disorders. 0.7657} & F1022: 0.894. Alcohol dependence with intoxication   &  F1721: 0.880. Nicotine dependence \\
 & F4310: 0.864. Post-traumatic stress disorder  & F11: 0.833. Opioid-related disorders    \\
 & F141: 0.826. Cocaine abuse &  F039: 0.819. Unspecified dementia  \\

\midrule
\multirow{4}{*}{VI: Nervous System Disorders. 0.7266} & G318: 0.838. Other specified degenerative diseases of the nervous system  &  G629: 0.832. Polyneuropathy, unspecified   \\
 & G609: 0.827. Hereditary and idiopathic neuropathy & G893: 0.822. Neoplasm-related pain (acute) (chronic)   \\
 & G931: 0.813. Anoxic brain damage   & G309: 0.806. Alzheimer's disease  \\
 & G20: 0.803. Parkinson's disease &    \\

\midrule
\multirow{1}{*}{VIII: Ear and Mastoid Disorders. 0.7501} &  H9190: 0.809. Unspecified hearing loss, unspecified ear  &        \\

\midrule
\multirow{9}{*}{IX: Circulatory System Diseases. 0.8428} & \textbf{I210: 0.986. ST elevation (STEMI) myocardial infarction of anterior wall}    &   \textbf{I314: 0.979. Cardiac tamponade}      \\
 &  \textbf{I447: 0.976. Left bundle-branch block}  &  \textbf{I481: 0.966. Persistent atrial fibrillation}   \\
 & \textbf{I451: 0.964. Other right bundle-branch block}   &  \textbf{I255: 0.964. Ischemic cardiomyopathy} \\
 & \textbf{I132: 0.949. Hypertensive heart failure and chronic kidney disease or end-stage renal disease} & \textbf{I078: 0.948. Other rheumatic tricuspid valve diseases}    \\
 & \textbf{I081: 0.944. Rheumatic disorders of both mitral and tricuspid valves} & \textbf{I2789: 0.943. Other specified pulmonary heart diseases} \\
 & \textbf{I5043: 0.940. Acute on chronic combined systolic and diastolic heart failure} & \textbf{I7025: 0.927. Atherosclerosis of native arteries with ulceration}  \\
\midrule

\multirow{3}{*}{X: Respiratory System Diseases. 0.8038} & \textbf{J9621: 0.951. Acute and chronic respiratory failure with hypoxia} & \textbf{J94: 0.917. Other pleural conditions}    \\
 & \textbf{J80: 0.905. Acute respiratory distress syndrome} & \textbf{J910: 0.904. Malignant pleural effusion}  \\
 & J848: 0.883. Interstitial pulmonary diseases & J90: 0.879. Pleural effusion, not elsewhere classified  \\

\midrule
\multirow{3}{*}{XI: Digestive System Diseases. 0.7409} & \textbf{K7031: 0.973. Alcoholic cirrhosis of the liver with ascites}  &  \textbf{K762: 0.948. Central hemorrhagic necrosis of liver} \\
 & \textbf{K7290: 0.947. Hepatic failure without coma}  & \textbf{K3189: 0.921. Other diseases of stomach and duodenum}   \\
 & K65: 0.853. Peritonitis  &  K830: 0.848. Cholangitis \\

\midrule
\multirow{2}{*}{XII: Skin and Subcutaneous Tissue Diseases. 0.7903} & L891: 0.874. Pressure ulcer of back   &  L9740: 0.870. Non-pressure chronic ulcer     \\
 & L0312: 0.851. Acute lymphangitis of a limb part &     \\

\midrule
\multirow{4}{*}{XIII: Musculoskeletal and Connective Tissue Diseases. 0.6903} & M129: 0.847. Arthropathy, unspecified    &  M949: 0.819. Disorder of cartilage       \\
 & M858: 0.805. Other specified disorders of bone density and structure   & M353: 0.805. Polymyalgia rheumatica    \\
 & M10: 0.803. Gout   & M321: 0.801. Systemic lupus erythematosus with organ or system involvement    \\

\midrule
\multirow{4}{*}{XIV: Genitourinary System Diseases. 0.7588} & N186: 0.887. End stage renal disease  &  N08: 0.878. Glomerular disorders in diseases classified elsewhere    \\
 & N9982: 0.857. Postprocedural hemorrhage of a genitourinary system organ  &  N170: 0.852. Acute kidney failure with tubular necrosis \\
 & N40: 0.811. Benign prostatic hyperplasia  &    \\  
 
\midrule
\multirow{3}{*}{XVIII: Symptoms and Clinical Findings. 0.7162} &  \textbf{R570: 0.931. Cardiogenic shock}  &  \textbf{R64: 0.900. Cachexia} \\
 & R18: 0.887. Ascites  &  R6521: 0.887. Severe sepsis with septic shock   \\
 & R34: 0.882. Anuria and oliguria   &  R000: 0.849. Tachycardia, unspecified  \\

\midrule
\multirow{5}{*}{XIX: Injuries and External Causes. 0.7765} &  \textbf{T8612: 0.944. Kidney transplant failure}  & T8285: 0.898. Stenosis due to cardiac and vascular prosthetic devices \\
 &  T380: 0.898. Poisoning by, adverse effect of and underdosing of glucocorticoids and synthetic analogues  &     \\
 &  T4551: 0.880. Poisoning by, adverse effect of and underdosing of anticoagulants  & T36: 0.872. Poisoning by, adverse effect of and underdosing of systemic antibiotics  \\
 
\midrule
\multirow{7}{*}{XX: External Causes of Morbidity and Mortality. 0.798} & \textbf{V422: 0.955. Person on outside of car injured in collision with two- or three-wheeled motor vehicle in nontraffic accident}   &   \textbf{V850: 0.949. Driver of special construction vehicle injured in traffic accident}      \\
 & \textbf{V433: 0.9281. Unspecified car occupant injured in collision with car, pick-up truck or van in nontraffic accident}   &  \textbf{V667: 0.919. Person on outside of heavy transport vehicle injured in collision with other nonmotor vehicle in traffic accident}   \\
 &  \textbf{V462: 0.900. Person on outside of car injured in collision with other nonmotor vehicle in nontraffic accident}  &  Y830: 0.8799. Surgical operation with transplant of whole organ as the cause of abnormal reaction of the patient, or of later complication, without mention of misadventure at the time of the procedure  \\

\midrule
\multirow{4}{*}{XXI: Health Status and Services. 0.7928} & \textbf{Z998: 0.946. Dependence on other enabling machines and devices}   &  \textbf{Z681: 0.944. Body mass index (BMI) 19.9 or less, adult}   \\
 & \textbf{Z4502: 0.941. Encounter for adjustment and management of automatic implantable cardiac defibrillator}  & \textbf{Z950: 0.940. Presence of cardiac pacemaker}   \\
 &  \textbf{Z590: 0.923. Homelessness}  & \textbf{Z515: 0.907. Encounter for palliative care}  \\

\bottomrule
\end{tabular}
}

\caption{T(ED2ALL)-E(ED2ALL) model (extended version of \Cref{tab:big_all_full}: Best-performing individual statements organized according to ICD chapters underscoring the breadth of accurately predictable statements. The table shows the six best-performing individual statements per ICD chapter (10 for chapter IX (Circulatory system diseases)), where we show only AUROC scores above 0.8. To showcase the breadth of reliably predictable statements, we list only the best-performing statement per 3-digit ICD code. Statements with an AUROC score of 0.9 or higher are marked in boldface.}

\label{tab:big_all_full}
\end{table}

\subsection{T(ED2ALL)-E(ED2ALL) Model: Statements with top label AUROCs}
\Cref{tab:big_all_full} shows as extended version of \Cref{tab:big_all}, i.e., the best-performing ICD statements.In \Cref{tab:auroc09} we summarize very accurately predictable statements (AUROC>0.9) aggregated at a 3-digit level. In \Cref{tab:auroc08}-\Cref{tab:auroc08part4} we summarize accurately predictable statements (AUROC>0.8). Finally in \Cref{tab:auroc07x}-\Cref{tab:auroc07xpart3}, we show challenging statements (AUROC<0.7).

\begin{table}[ht]
    \centering
    \scalebox{0.8}{
    \begin{tabular}{llllp{8cm}}
    \toprule
Chapter & Code & Coverage & Prevalence & Description\\
\midrule
II & C79 & 1/8 &0.015 & Secondary malignant neoplasm of other and unspecified sites\\
II & C92 & 1/2 &0.002 & Myeloid leukemia\\
\midrule
III & \textbf{D65} & 1/1 &0.001 & Disseminated intravascular coagulation [defibrination syndrome]\\
\midrule
IV & E11 & 2/18 &0.217 & Type 2 diabetes mellitus\\
IV & E66 & 2/4 &0.062 & Overweight and obesity\\
\midrule
IX & I25 & 3/10 &0.162 & Chronic ischemic heart disease\\
IX & \textbf{I48} & 6/7 &0.152 & Atrial fibrillation and flutter\\
IX & I50 & 9/15 &0.142 & Heart failure\\
IX & I27 & 3/5 &0.037 & Other pulmonary heart diseases\\
IX & \textbf{I21} & 6/8 &0.036 & Acute myocardial infarction\\
IX & \textbf{I13} & 3/3 &0.025 & Hypertensive heart and chronic kidney disease\\
IX & \textbf{I11} & 2/2 &0.022 & Hypertensive heart disease\\
IX & I34 & 1/3 &0.020 & Nonrheumatic mitral valve disorders\\
IX & I42 & 1/3 &0.018 & Cardiomyopathy\\
IX & \textbf{I44} & 5/5 &0.015 & Atrioventricular and left bundle-branch block\\
IX & I45 & 2/5 &0.013 & Other conduction disorders\\
IX & I70 & 1/5 &0.008 & Atherosclerosis\\
IX & I08 & 1/3 &0.008 & Multiple valve diseases\\
IX & I31 & 2/4 &0.007 & Other diseases of pericardium\\
IX & I07 & 1/2 &0.005 & Rheumatic tricuspid valve diseases\\
\midrule
X & J96 & 2/10 &0.049 & Respiratory failure, not elsewhere classified\\
X & J91 & 1/3 &0.011 & Pleural effusion in conditions classified elsewhere\\
X & \textbf{J94} & 2/2 &0.001 & Other pleural conditions\\
X & \textbf{J80} & 1/1 &0.000 & Acute respiratory distress syndrome\\
\midrule
XI & K76 & 2/6 &0.023 & Other diseases of liver\\
XI & K31 & 1/4 &0.022 & Other diseases of stomach and duodenum\\
XI & \textbf{K70} & 4/5 &0.016 & Alcoholic liver disease\\
XI & \textbf{K72} & 6/6 &0.006 & Hepatic failure, not elsewhere classified\\
\midrule
XVIII & R57 & 2/4 &0.011 & Shock, not elsewhere classified\\
\midrule
XIX & T86 & 2/3 &0.004 & Complications of transplanted organs and tissue\\
\midrule
XX & V43 & 1/4 &0.014 & Car occupant injured in collision with car, pick-up truck or van\\
XX & V46 & 1/2 &0.011 & Car occupant injured in collision with other nonmotor vehicle\\
XX & V42 & 2/3 &0.007 & Car occupant injured in collision with two- or three-wheeled motor vehicle\\
XX & \textbf{V66} & 2/2 &0.006 & Occupant of heavy transport vehicle injured in collision with other nonmotor vehicle\\
XX & \textbf{V85} & 2/2 &0.005 & Occupant of special construction vehicle injured in transport accident\\
\midrule
XXI & Z95 & 5/7 &0.084 & Presence of cardiac and vascular implants and grafts\\
XXI & Z68 & 2/8 &0.060 & Body mass index (BMI)\\
XXI & Z99 & 2/5 &0.033 & Dependence on enabling machines and devices, not elsewhere classified\\
XXI & Z51 & 1/2 &0.010 & Encounter for other aftercare and medical care\\
XXI & \textbf{Z45} & 4/4 &0.005 & Encounter for adjustment and management of implanted device\\
XXI & Z59 & 1/2 &0.005 & Problems related to housing and economic circumstances\\    
\bottomrule
    \end{tabular}
    }
    \caption{T(ED2ALL)-E(ED2ALL) model: Statements with AUROCs larger than 0.9. The statements are sorted by chapter and prevalence in the dataset. Coverage refers to the fraction of codes within this 3-digit category that exceed the specified threshold of 0.9. Categories in boldface correspond to categories with a coverage of 75\% or higher, i.e., a situation where the category and 75\% of the subcategories are covered with AUROC scores above 0.9.}
    \label{tab:auroc09}
\end{table}

\begin{table}[ht]
    \centering
    \scalebox{0.8}{
    \begin{tabular}{llllp{8cm}}
    \toprule
Chapter & Code & Coverage & Prevalence & Description\\
\midrule
I & \textbf{A41} & 7/7 &0.038 & Other sepsis\\
I & B96 & 2/9 &0.036 & Other bacterial agents as the cause of diseases classified elsewhere\\
I & A04 & 1/2 &0.014 & Other bacterial intestinal infections\\
I & B37 & 2/4 &0.011 & Candidiasis\\
I & \textbf{A40} & 1/1 &0.002 & Streptococcal sepsis\\
\midrule
II & \textbf{C79} & 7/8 &0.015 & Secondary malignant neoplasm of other and unspecified sites\\
II & C78 & 3/5 &0.014 & Secondary malignant neoplasm of respiratory and digestive organs\\
II & \textbf{C34} & 3/4 &0.012 & Malignant neoplasm of bronchus and lung\\
II & \textbf{C61} & 1/1 &0.005 & Malignant neoplasm of prostate\\
II & \textbf{C25} & 1/1 &0.005 & Malignant neoplasm of pancreas\\
II & \textbf{D46} & 2/2 &0.003 & Myelodysplastic syndromes\\
II & C85 & 2/3 &0.002 & Other specified and unspecified types of non-Hodgkin lymphoma\\
II & \textbf{C92} & 2/2 &0.002 & Myeloid leukemia\\
\midrule
III & \textbf{D63} & 3/4 &0.041 & Anemia in chronic diseases classified elsewhere\\
III & D68 & 2/4 &0.024 & Other coagulation defects\\
III & \textbf{D61} & 3/3 &0.012 & Other aplastic anemias and other bone marrow failure syndromes\\
III & \textbf{D65} & 1/1 &0.001 & Disseminated intravascular coagulation [defibrination syndrome]\\
\midrule
IV & E78 & 1/4 &0.285 & Disorders of lipoprotein metabolism and other lipidemias\\
IV & E11 & 12/18 &0.217 & Type 2 diabetes mellitus\\
IV & E87 & 5/11 &0.145 & Other disorders of fluid, electrolyte and acid-base balance\\
IV & \textbf{E66} & 3/4 &0.062 & Overweight and obesity\\
IV & E83 & 3/9 &0.029 & Disorders of mineral metabolism\\
IV & E88 & 2/5 &0.025 & Other and unspecified metabolic disorders\\
IV & \textbf{E10} & 6/7 &0.018 & Type 1 diabetes mellitus\\
IV & \textbf{E13} & 3/3 &0.014 & Other specified diabetes mellitus\\
IV & \textbf{E43} & 1/1 &0.012 & Unspecified severe protein-calorie malnutrition\\
IV & \textbf{E46} & 1/1 &0.011 & Unspecified protein-calorie malnutrition\\
IV & \textbf{E16} & 2/2 &0.004 & Other disorders of pancreatic internal secretion\\
IV & E44 & 1/2 &0.004 & Protein-calorie malnutrition of moderate and mild degree\\
\midrule
V & F17 & 1/4 &0.076 & Nicotine dependence\\
V & F10 & 3/8 &0.059 & Alcohol related disorders\\
V & \textbf{F03} & 3/3 &0.027 & Unspecified dementia\\
V & F43 & 2/5 &0.017 & Reaction to severe stress, and adjustment disorders\\
V & \textbf{F02} & 3/3 &0.014 & Dementia in other diseases classified elsewhere\\
V & \textbf{F11} & 5/5 &0.013 & Opioid-related disorders\\
V & \textbf{F14} & 3/3 &0.012 & Cocaine related disorders\\
V & \textbf{F01} & 3/3 &0.005 & Vascular dementia\\
V & F34 & 1/2 &0.002 & Persistent mood [affective] disorders\\
\midrule
VI & G89 & 1/5 &0.044 & Pain, not elsewhere classified\\
VI & G93 & 1/7 &0.033 & Other disorders of brain\\
VI & \textbf{G30} & 2/2 &0.010 & Alzheimer's disease\\
VI & G60 & 1/2 &0.008 & Hereditary and idiopathic neuropathy\\
VI & \textbf{G20} & 1/1 &0.008 & Parkinson's disease\\
VI & \textbf{G62} & 2/2 &0.007 & Other and unspecified polyneuropathies\\
VI & \textbf{G31} & 2/2 &0.007 & Other degenerative diseases of nervous system, not elsewhere classified\\
    \bottomrule
    \end{tabular}
    }
    \caption{T(ED2ALL)-E(ED2ALL) model: Statements with AUROCs larger than 0.8 (Part 1). The statements are sorted by chapter and prevalence in the dataset. Coverage refers to the fraction of codes within this 3-digit category that exceed the specified threshold of 0.8. Categories in boldface correspond to categories with a coverage of 75\% or higher, i.e., a situation where the category and 75\% of the subcategories are covered with AUROC scores above 0.8.}
    \label{tab:auroc08}
\end{table}

\begin{table}[ht]
    \centering
    \scalebox{0.8}{
    \begin{tabular}{llllp{8cm}}
    \toprule
     Chapter & Code & Coverage & Prevalence & Description\\
   \midrule
VIII & H91 & 1/3 &0.008 & Other and unspecified hearing loss\\
\midrule
IX & \textbf{I25} & 9/10 &0.162 & Chronic ischemic heart disease\\
IX & \textbf{I48} & 7/7 &0.152 & Atrial fibrillation and flutter\\
IX & \textbf{I50} & 15/15 &0.142 & Heart failure\\
IX & I12 & 2/3 &0.090 & Hypertensive chronic kidney disease\\
IX & I95 & 2/7 &0.055 & Hypotension\\
IX & \textbf{I27} & 5/5 &0.037 & Other pulmonary heart diseases\\
IX & \textbf{I21} & 8/8 &0.036 & Acute myocardial infarction\\
IX & \textbf{I13} & 3/3 &0.025 & Hypertensive heart and chronic kidney disease\\
IX & \textbf{I11} & 2/2 &0.022 & Hypertensive heart disease\\
IX & \textbf{I34} & 3/3 &0.020 & Nonrheumatic mitral valve disorders\\
IX & \textbf{I42} & 3/3 &0.018 & Cardiomyopathy\\
IX & \textbf{I47} & 3/3 &0.018 & Paroxysmal tachycardia\\
IX & \textbf{I44} & 5/5 &0.015 & Atrioventricular and left bundle-branch block\\
IX & I69 & 4/6 &0.015 & Sequelae of cerebrovascular disease\\
IX & I82 & 1/5 &0.014 & Other venous embolism and thrombosis\\
IX & I63 & 1/5 &0.013 & Cerebral infarction\\
IX & I35 & 2/3 &0.013 & Nonrheumatic aortic valve disorders\\
IX & \textbf{I45} & 5/5 &0.013 & Other conduction disorders\\
IX & I67 & 1/3 &0.012 & Other cerebrovascular diseases\\
IX & \textbf{I24} & 2/2 &0.009 & Other acute ischemic heart diseases\\
IX & I20 & 2/3 &0.008 & Angina pectoris\\
IX & \textbf{I70} & 4/5 &0.008 & Atherosclerosis\\
IX & \textbf{I08} & 3/3 &0.008 & Multiple valve diseases\\
IX & I51 & 1/2 &0.007 & Complications and ill-defined descriptions of heart disease\\
IX & \textbf{I31} & 4/4 &0.007 & Other diseases of pericardium\\
IX & \textbf{I85} & 3/3 &0.006 & Esophageal varices\\
IX & I65 & 1/3 &0.006 & Occlusion and stenosis of precerebral arteries, not resulting in cerebral infarction\\
IX & \textbf{I07} & 2/2 &0.005 & Rheumatic tricuspid valve diseases\\
IX & \textbf{I46} & 2/2 &0.004 & Cardiac arrest\\
IX & \textbf{I16} & 1/1 &0.003 & Hypertensive crisis\\
\midrule
X & J45 & 1/4 &0.068 & Asthma\\
X & J44 & 2/3 &0.066 & Other chronic obstructive pulmonary disease\\
X & \textbf{J96} & 10/10 &0.049 & Respiratory failure, not elsewhere classified\\
X & J98 & 3/7 &0.018 & Other respiratory disorders\\
X & \textbf{J91} & 3/3 &0.011 & Pleural effusion in conditions classified elsewhere\\
X & \textbf{J84} & 3/3 &0.008 & Other interstitial pulmonary diseases\\
X & J15 & 1/4 &0.007 & Bacterial pneumonia, not elsewhere classified\\
X & J95 & 2/4 &0.006 & Intraoperative and postprocedural complications and disorders of respiratory system, not elsewhere classified\\
X & \textbf{J90} & 1/1 &0.006 & Pleural effusion, not elsewhere classified\\
X & \textbf{J47} & 1/1 &0.003 & Bronchiectasis\\
X & \textbf{J94} & 2/2 &0.001 & Other pleural conditions\\
X & \textbf{J80} & 1/1 &0.000 & Acute respiratory distress syndrome\\
\midrule
XI & \textbf{K76} & 6/6 &0.023 & Other diseases of liver\\
XI & \textbf{K31} & 4/4 &0.022 & Other diseases of stomach and duodenum\\
XI & \textbf{K70} & 5/5 &0.016 & Alcoholic liver disease\\
XI & K83 & 2/3 &0.010 & Other diseases of biliary tract\\
XI & \textbf{K72} & 6/6 &0.006 & Hepatic failure, not elsewhere classified\\
XI & K65 & 1/2 &0.004 & Peritonitis\\
    \bottomrule
    \end{tabular}
    }
    \caption{T(ED2ALL)-E(ED2ALL) model: Statements with AUROCs larger than 0.8 (Part 2). The statements are sorted by chapter and prevalence in the dataset. Coverage refers to the fraction of codes within this 3-digit category that exceed the specified threshold of 0.8. Categories in boldface correspond to categories with a coverage of 75\% or higher, i.e., a situation where the category and 75\% of the subcategories are covered with AUROC scores above 0.8.}
    \label{tab:auroc08part2}
\end{table}

\begin{table}[ht]
    \centering
    \scalebox{0.8}{
    \begin{tabular}{llllp{8cm}}
    \toprule
    Chapter & Code & Coverage & Prevalence & Description\\
\midrule
XII & L03 & 2/5 &0.016 & Cellulitis and acute lymphangitis\\
XII & \textbf{L89} & 3/3 &0.013 & Pressure ulcer\\
XII & \textbf{L97} & 4/5 &0.010 & Non-pressure chronic ulcer of lower limb, not elsewhere classified\\
\midrule
XIII & \textbf{M10} & 2/2 &0.034 & Gout\\
XIII & M94 & 1/2 &0.011 & Other disorders of cartilage\\
XIII & M35 & 1/2 &0.005 & Other systemic involvement of connective tissue\\
XIII & \textbf{M85} & 2/2 &0.004 & Other disorders of bone density and structure\\
XIII & M32 & 1/2 &0.003 & Systemic lupus erythematosus (SLE)\\
XIII & \textbf{M12} & 2/2 &0.002 & Other and unspecified arthropathy\\
\midrule
XIV & N18 & 4/7 &0.131 & Chronic kidney disease (CKD)\\
XIV & N17 & 1/3 &0.121 & Acute kidney failure\\
XIV & \textbf{N40} & 3/3 &0.044 & Benign prostatic hyperplasia\\
XIV & \textbf{N08} & 1/1 &0.006 & Glomerular disorders in diseases classified elsewhere\\
XIV & N99 & 2/3 &0.004 & Intraoperative and postprocedural complications and disorders of genitourinary system, not elsewhere classified\\
\midrule
XVII & \textbf{Q23} & 1/1 &0.001 & Congenital malformations of aortic and mitral valves\\
\midrule
XVIII & R00 & 1/4 &0.070 & Abnormalities of heart beat\\
XVIII & R53 & 1/4 &0.030 & Malaise and fatigue\\
XVIII & \textbf{R65} & 4/4 &0.027 & Symptoms and signs specifically associated with systemic inflammation and infection\\
XVIII & R41 & 1/4 &0.025 & Other symptoms and signs involving cognitive functions and awareness\\
XVIII & R33 & 1/3 &0.023 & Retention of urine\\
XVIII & R13 & 1/4 &0.021 & Aphagia and dysphagia\\
XVIII & \textbf{R57} & 4/4 &0.011 & Shock, not elsewhere classified\\
XVIII & R40 & 2/3 &0.010 & Somnolence, stupor and coma\\
XVIII & \textbf{R18} & 2/2 &0.010 & Ascites\\
XVIII & \textbf{R64} & 1/1 &0.007 & Cachexia\\
XVIII & \textbf{R34} & 1/1 &0.002 & Anuria and oliguria\\
\midrule
XIX & T82 & 4/7 &0.011 & Complications of cardiac and vascular prosthetic devices, implants and grafts\\
XIX & \textbf{T45} & 4/4 &0.011 & Poisoning by, adverse effect of and underdosing of primarily systemic and hematological agents, not elsewhere classified\\
XIX & S72 & 1/2 &0.010 & Fracture of femur\\
XIX & T40 & 1/2 &0.008 & Poisoning by, adverse effect of and underdosing of narcotics and psychodysleptics [hallucinogens]\\
XIX & \textbf{T50} & 1/1 &0.008 & Poisoning by, adverse effect of and underdosing of diuretics and other and unspecified drugs, medicaments and biological substances\\
XIX & T81 & 2/5 &0.007 & Complications of procedures, not elsewhere classified\\
XIX & \textbf{T38} & 2/2 &0.005 & Poisoning by, adverse effect of and underdosing of hormones and their synthetic substitutes and antagonists, not elsewhere classified\\
XIX & \textbf{T86} & 3/3 &0.004 & Complications of transplanted organs and tissue\\
XIX & \textbf{T36} & 1/1 &0.004 & Poisoning by, adverse effect of and underdosing of systemic antibiotics\\
    \bottomrule
    \end{tabular}
    }
    \caption{T(ED2ALL)-E(ED2ALL) model: Statements with AUROCs larger than 0.8 (Part 3). The statements are sorted by chapter and prevalence in the dataset. Coverage refers to the fraction of codes within this 3-digit category that exceed the specified threshold of 0.8. Categories in boldface correspond to categories with a coverage of 75\% or higher, i.e., a situation where the category and 75\% of the subcategories are covered with AUROC scores above 0.8.}
    \label{tab:auroc08part3}
\end{table}

\begin{table}[ht]
    \centering
    \scalebox{0.8}{
    \begin{tabular}{llllp{8cm}}
    \toprule
    Chapter & Code & Coverage & Prevalence & Description\\
\midrule
XX & Y92 & 4/12 &0.122 & Place of occurrence of the external cause\\
XX & Y83 & 1/6 &0.031 & Surgical operation and other surgical procedures as the cause of abnormal reaction of the patient, or of later complication, without mention of misadventure at the time of the procedure\\
XX & W18 & 2/5 &0.029 & Other slipping, tripping and stumbling and falls\\
XX & Y84 & 1/4 &0.024 & Other medical procedures as the cause of abnormal reaction of the patient, or of later complication, without mention of misadventure at the time of the procedure\\
XX & \textbf{V43} & 4/4 &0.014 & Car occupant injured in collision with car, pick-up truck or van\\
XX & \textbf{V17} & 2/2 &0.013 & Pedal cycle rider injured in collision with fixed or stationary object\\
XX & \textbf{V15} & 2/2 &0.012 & Pedal cycle rider injured in collision with railway train or railway vehicle\\
XX & \textbf{V46} & 2/2 &0.011 & Car occupant injured in collision with other nonmotor vehicle\\
XX & \textbf{V10} & 2/2 &0.010 & Pedal cycle rider injured in collision with pedestrian or animal\\
XX & \textbf{V42} & 3/3 &0.007 & Car occupant injured in collision with two- or three-wheeled motor vehicle\\
XX & \textbf{V66} & 2/2 &0.006 & Occupant of heavy transport vehicle injured in collision with other nonmotor vehicle \\
XX & \textbf{V44} & 1/1 &0.005 & Car occupant injured in collision with heavy transport vehicle or bus\\
XX & \textbf{V60} & 2/2 &0.005 & Occupant of heavy transport vehicle injured in collision with pedestrian or animal\\
XX & \textbf{V85} & 2/2 &0.005 & Occupant of special construction vehicle injured in transport accident\\
XX & \textbf{V45} & 2/2 &0.002 & Car occupant injured in collision with railway train or railway vehicle\\
XX & \textbf{V64} & 1/1 &0.000 & Occupant of heavy transport vehicle injured in collision with heavy transport vehicle or bus\\
\midrule
XXI & Z79 & 6/11 &0.237 & Long term (current) drug therapy\\
XXI & Z85 & 2/15 &0.093 & Personal history of malignant neoplasm\\
XXI & \textbf{Z95} & 7/7 &0.084 & Presence of cardiac and vascular implants and grafts\\
XXI & \textbf{Z66} & 1/1 &0.062 & Do not resuscitate\\
XXI & \textbf{Z68} & 8/8 &0.060 & Body mass index [BMI]\\
XXI & Z91 & 1/6 &0.049 & Personal risk factors, not elsewhere classified\\
XXI & Z98 & 3/6 &0.045 & Other postprocedural states\\
XXI & \textbf{Z99} & 5/5 &0.033 & Dependence on enabling machines and devices, not elsewhere classified\\
XXI & Z90 & 2/9 &0.028 & Acquired absence of organs, not elsewhere classified\\
XXI & Z96 & 1/4 &0.027 & Presence of other functional implants\\
XXI & Z92 & 1/4 &0.018 & Personal history of medical treatment\\
XXI & \textbf{Z51} & 2/2 &0.010 & Encounter for other aftercare and medical care\\
XXI & \textbf{Z89} & 5/5 &0.009 & Acquired absence of limb\\
XXI & \textbf{Z72} & 2/2 &0.007 & Problems related to lifestyle\\
XXI & Z94 & 2/3 &0.006 & Transplanted organ and tissue status\\
XXI & \textbf{Z00} & 2/2 &0.005 & Encounter for general examination without complaint, suspected or reported diagnosis\\
XXI & \textbf{Z45} & 4/4 &0.005 & Encounter for adjustment and management of implanted device\\
XXI & \textbf{Z59} & 2/2 &0.005 & Problems related to housing and economic circumstances\\
XXI & \textbf{Z93} & 1/1 &0.005 & Artificial opening status\\
XXI & \textbf{Z76} & 3/3 &0.004 & Persons encountering health services in other circumstances\\    
\bottomrule
    \end{tabular}
    }
    \caption{T(ED2ALL)-E(ED2ALL) model: Statements with AUROCs larger than 0.8 (Part 4). The statements are sorted by chapter and prevalence in the dataset. Coverage refers to the fraction of codes within this 3-digit category that exceed the specified threshold of 0.8. Categories in boldface correspond to categories with a coverage of 75\% or higher, i.e., a situation where the category and 75\% of the subcategories are covered with AUROC scores above 0.8.}
    \label{tab:auroc08part4}
\end{table}

\begin{table}[ht]
    \centering
    \scalebox{0.8}{
    \begin{tabular}{llllp{8cm}}
    \toprule
    Chapter & Code & Coverage & Prevalence & Description\\
\midrule
I & B95 & 1/6 &0.017 & Streptococcus, Staphylococcus, and Enterococcus as the cause of diseases classified elsewhere\\
I & \textbf{B97} & 1/1 &0.004 & Viral agents as the cause of diseases classified elsewhere\\
I & \textbf{B00} & 1/1 &0.001 & Herpesviral [herpes simplex] infections\\
\midrule
II & D47 & 1/3 &0.008 & Other neoplasms of uncertain behavior of lymphoid, hematopoietic and related tissue\\
\midrule
III & \textbf{D72} & 5/5 &0.035 & Other disorders of white blood cells\\
III & D50 & 2/3 &0.027 & Iron deficiency anemia\\
III & D70 & 1/2 &0.004 & Neutropenia\\
III & \textbf{D86} & 1/1 &0.002 & Sarcoidosis\\
III & \textbf{D75} & 1/1 &0.001 & Other and unspecified diseases of blood and blood-forming organs\\
\midrule
IV & E78 & 1/4 &0.285 & Disorders of lipoprotein metabolism and other lipidemias\\
IV & \textbf{E86} & 3/4 &0.057 & Volume depletion\\
IV & \textbf{E55} & 2/2 &0.011 & Vitamin D deficiency\\
IV & \textbf{E04} & 2/2 &0.007 & Other nontoxic goiter\\
IV & E53 & 1/2 &0.007 & Deficiency of other B group vitamins\\
IV & \textbf{E89} & 2/2 &0.006 & Postprocedural endocrine and metabolic complications and disorders, not elsewhere classified\\
IV & \textbf{E05} & 1/1 &0.006 & Thyrotoxicosis [hyperthyroidism]\\
IV & \textbf{E21} & 2/2 &0.004 & Hyperparathyroidism and other disorders of parathyroid gland\\
\midrule
V & \textbf{F32} & 2/2 &0.116 & Major depressive disorder, single episode\\
V & \textbf{F41} & 3/3 &0.090 & Other anxiety disorders\\
V & \textbf{F31} & 2/2 &0.023 & Bipolar disorder\\
V & F43 & 1/5 &0.017 & Reaction to severe stress, and adjustment disorders\\
V & \textbf{F25} & 2/2 &0.006 & Schizoaffective disorders\\
V & F06 & 1/2 &0.001 & Other mental disorders due to known physiological condition\\
\midrule
VI & G47 & 2/6 &0.080 & Sleep disorders\\
VI & G89 & 3/5 &0.044 & Pain, not elsewhere classified\\
VI & G93 & 1/7 &0.033 & Other disorders of brain\\
VI & \textbf{G40} & 3/3 &0.026 & Epilepsy and recurrent seizures\\
VI & \textbf{G43} & 3/3 &0.015 & Migraine\\
VI & \textbf{G25} & 3/3 &0.010 & Other extrapyramidal and movement disorders\\
VI & \textbf{G81} & 2/2 &0.008 & Hemiplegia and hemiparesis\\
VI & \textbf{G91} & 1/1 &0.003 & Hydrocephalus\\
\midrule
VII & \textbf{H53} & 1/1 &0.010 & Visual disturbances\\
VII & \textbf{H26} & 2/2 &0.003 & Other cataract\\
\midrule
VIII & \textbf{H81} & 1/1 &0.007 & Disorders of vestibular function\\
\midrule
IX & I95 & 2/7 &0.055 & Hypotension\\
IX & I67 & 1/3 &0.012 & Other cerebrovascular diseases\\
IX & \textbf{I77} & 1/1 &0.005 & Other disorders of arteries and arterioles\\
IX & I80 & 1/2 &0.001 & Phlebitis and thrombophlebitis\\
    \bottomrule
    \end{tabular}
    }
    \caption{T(ED2ALL)-E(ED2ALL) model: Statements with AUROCs smaller than 0.7 (Part 1). The statements are sorted by chapter and prevalence in the dataset. Coverage refers to the fraction of codes within this 3-digit category that stay below the specified threshold of 0.7. Categories in boldface correspond to categories with a coverage of 75\% or higher, i.e., a situation where the category and 75\% of the subcategories are covered with AUROC scores below 0.7.}
    \label{tab:auroc07x}
\end{table}

\begin{table}[ht]
    \centering
    \scalebox{0.8}{
    \begin{tabular}{llllp{8cm}}
    \toprule
    Chapter & Code & Coverage & Prevalence & Description\\
\midrule
X & J45 & 2/4 &0.068 & Asthma\\
X & \textbf{J06} & 2/2 &0.010 & Acute upper respiratory infections of multiple and unspecified sites\\
X & \textbf{J38} & 1/1 &0.002 & Diseases of vocal cords and larynx, not elsewhere classified\\
\midrule
XI & \textbf{K21} & 2/2 &0.142 & Gastro-esophageal reflux disease\\
XI & \textbf{K59} & 4/4 &0.042 & Other functional intestinal disorders\\
XI & K92 & 1/4 &0.021 & Other diseases of digestive system\\
XI & \textbf{K22} & 3/3 &0.017 & Other diseases of esophagus\\
XI & K29 & 1/3 &0.014 & Gastritis and duodenitis\\
XI & \textbf{K56} & 3/4 &0.012 & Paralytic ileus and intestinal obstruction without hernia\\
XI & \textbf{K86} & 2/2 &0.010 & Other diseases of pancreas\\
XI & \textbf{K80} & 3/3 &0.010 & Cholelithiasis\\
XI & \textbf{K52} & 2/2 &0.009 & Other and unspecified noninfective gastroenteritis and colitis\\
XI & \textbf{K58} & 2/2 &0.008 & Irritable bowel syndrome\\
XI & \textbf{K85} & 2/2 &0.008 & Acute pancreatitis\\
XI & \textbf{K50} & 3/3 &0.007 & Crohn's disease [regional enteritis]\\
XI & \textbf{K62} & 1/1 &0.006 & Other diseases of anus and rectum\\
XI & \textbf{K81} & 1/1 &0.002 & Cholecystitis\\
XI & \textbf{K43} & 1/1 &0.002 & Ventral hernia\\
XI & \textbf{K51} & 1/1 &0.002 & Ulcerative colitis\\
XI & \textbf{K66} & 2/2 &0.002 & Other disorders of peritoneum\\
XI & \textbf{K12} & 2/2 &0.001 & Stomatitis and related lesions\\
\midrule
XII & L27 & 1/2 &0.004 & Dermatitis due to substances taken internally\\
XII & \textbf{L98} & 1/1 &0.003 & Other disorders of skin and subcutaneous tissue, not elsewhere classified\\
\midrule
XIII & \textbf{M54} & 4/4 &0.049 & Dorsalgia\\
XIII & \textbf{M79} & 6/6 &0.036 & Other and unspecified soft tissue disorders, not elsewhere classified\\
XIII & \textbf{M25} & 5/5 &0.024 & Other joint disorder, not elsewhere classified\\
XIII & \textbf{M06} & 2/2 &0.014 & Other rheumatoid arthritis\\
XIII & \textbf{M48} & 3/4 &0.013 & Other spondylopathies\\
XIII & M62 & 2/3 &0.011 & Other disorders of muscle\\
XIII & M47 & 2/3 &0.009 & Spondylosis\\
XIII & M17 & 1/2 &0.007 & Osteoarthritis of knee\\
XIII & \textbf{M51} & 1/1 &0.004 & Thoracic, thoracolumbar, and lumbosacral intervertebral disc disorders\\
\midrule
XIV & \textbf{N28} & 2/2 &0.006 & Other disorders of kidney and ureter, not elsewhere classified\\
XIV & \textbf{N20} & 2/2 &0.004 & Calculus of kidney and ureter\\
XIV & \textbf{N31} & 2/2 &0.003 & Neuromuscular dysfunction of bladder, not elsewhere classified\\
\midrule
XVII & \textbf{Q21} & 2/2 &0.003 & Congenital malformations of cardiac septa\\
    \bottomrule
    \end{tabular}
    }
    \caption{T(ED2ALL)-E(ED2ALL) model: Statements with AUROCs smaller than 0.7 (Part 2). The statements are sorted by chapter and prevalence in the dataset. Coverage refers to the fraction of codes within this 3-digit category that stay below the specified threshold of 0.7. Categories in boldface correspond to categories with a coverage of 75\% or higher, i.e., a situation where the category and 75\% of the subcategories are covered with AUROC scores below 0.7.}
    \label{tab:auroc07xpart2}
\end{table}

\begin{table}[ht]
    \centering
    \scalebox{0.8}{
    \begin{tabular}{llllp{8cm}}
    \toprule
    Chapter & Code & Coverage & Prevalence & Description\\
\midrule
XVIII & \textbf{R07} & 3/4 &0.191 & Pain in throat and chest\\
XVIII & R00 & 1/4 &0.070 & Abnormalities of heart beat\\
XVIII & \textbf{R10} & 7/7 &0.068 & Abdominal and pelvic pain\\
XVIII & R06 & 3/6 &0.053 & Abnormalities of breathing\\
XVIII & \textbf{R55} & 1/1 &0.048 & Syncope and collapse\\
XVIII & \textbf{R11} & 4/4 &0.043 & Nausea and vomiting\\
XVIII & \textbf{R42} & 1/1 &0.043 & Dizziness and giddiness\\
XVIII & \textbf{R53} & 3/4 &0.030 & Malaise and fatigue\\
XVIII & \textbf{R51} & 1/1 &0.030 & Headache\\
XVIII & \textbf{R19} & 2/2 &0.029 & Other symptoms and signs involving the digestive system and abdomen\\
XVIII & \textbf{R41} & 3/4 &0.025 & Other symptoms and signs involving cognitive functions and awareness\\
XVIII & R79 & 1/4 &0.023 & Other abnormal findings of blood chemistry\\
XVIII & \textbf{R45} & 3/4 &0.022 & Symptoms and signs involving emotional state\\
XVIII & \textbf{R05} & 1/1 &0.020 & Cough\\
XVIII & R50 & 1/4 &0.020 & Fever of other and unknown origin\\
XVIII & R74 & 1/3 &0.019 & Abnormal serum enzyme levels\\
XVIII & \textbf{R47} & 4/4 &0.018 & Speech disturbances, not elsewhere classified\\
XVIII & R91 & 2/3 &0.016 & Abnormal findings on diagnostic imaging of lung\\
XVIII & \textbf{R29} & 3/4 &0.015 & Other symptoms and signs involving the nervous and musculoskeletal systems\\
XVIII & R20 & 1/2 &0.014 & Disturbances of skin sensation\\
XVIII & R60 & 1/2 &0.014 & Edema, not elsewhere classified\\
XVIII & R94 & 1/3 &0.011 & Abnormal results of function studies\\
XVIII & \textbf{R56} & 2/2 &0.010 & Convulsions, not elsewhere classified\\
XVIII & R04 & 1/2 &0.010 & Hemorrhage from respiratory passages\\
XVIII & R40 & 1/3 &0.010 & Somnolence, stupor and coma\\
XVIII & \textbf{R73} & 3/3 &0.008 & Elevated blood glucose level\\
XVIII & R78 & 1/3 &0.008 & Findings of drugs and other substances, not normally found in blood\\
XVIII & \textbf{R25} & 1/1 &0.005 & Abnormal involuntary movements\\
XVIII & \textbf{R21} & 1/1 &0.005 & Rash and other nonspecific skin eruption\\
XVIII & \textbf{R68} & 1/1 &0.004 & Other general symptoms and signs\\
XVIII & \textbf{R59} & 1/1 &0.003 & Enlarged lymph nodes\\
\midrule
XIX & S06 & 1/3 &0.014 & Intracranial injury\\
XIX & T82 & 1/7 &0.011 & Complications of cardiac and vascular prosthetic devices, implants and grafts\\
XIX & \textbf{S22} & 2/2 &0.010 & Fracture of rib(s), sternum and thoracic spine\\
XIX & \textbf{S82} & 1/1 &0.005 & Fracture of lower leg, including ankle\\
XIX & \textbf{S42} & 1/1 &0.005 & Fracture of shoulder and upper arm\\
XIX & \textbf{T84} & 1/1 &0.002 & Complications of internal orthopedic prosthetic devices, implants and grafts\\
\midrule
XX & Y92 & 3/12 &0.122 & Place of occurrence of the external cause\\
XX & Y83 & 4/6 &0.031 & Surgical operation and other surgical procedures as the cause of abnormal reaction of the patient, or of later complication, without mention of misadventure at the time of the procedure\\
XX & W18 & 2/5 &0.029 & Other slipping, tripping and stumbling and falls\\
XX & \textbf{X58} & 1/1 &0.011 & Exposure to other specified factors\\
\midrule
XXI & Z79 & 1/11 &0.237 & Long term (current) drug therapy\\
XXI & Z87 & 3/7 &0.163 & Personal history of other diseases and conditions\\
XXI & Z86 & 7/10 &0.099 & Personal history of certain other diseases\\
XXI & Z85 & 2/15 &0.093 & Personal history of malignant neoplasm\\
XXI & Z91 & 1/6 &0.049 & Personal risk factors, not elsewhere classified\\
XXI & Z98 & 1/6 &0.045 & Other postprocedural states\\
XXI & Z90 & 1/9 &0.028 & Acquired absence of organs, not elsewhere classified\\
XXI & \textbf{Z23} & 1/1 &0.025 & Encounter for immunization\\
XXI & \textbf{Z16} & 1/1 &0.003 & Resistance to antimicrobial drugs\\    
\bottomrule
    \end{tabular}
    }
    \caption{T(ED2ALL)-E(ED2ALL) model: Statements with AUROCs smaller than 0.7 (Part 3). The statements are sorted by chapter and prevalence in the dataset. Coverage refers to the fraction of codes within this 3-digit category that stay below the specified threshold of 0.7. Categories in boldface correspond to categories with a coverage of 75\% or higher, i.e., a situation where the category and 75\% of the subcategories are covered with AUROC scores below 0.7.}
    \label{tab:auroc07xpart3}
\end{table}

\subsection{T(ALL2ALL)-E(ALL2ALL) Model: Statements with top label AUROCs}
\Cref{tab:big_all_ALL} shows the analogue of \Cref{tab:big_all_full} for the T(ALL2ALL)-model, i.e., the best-performing ICD statements. In \Cref{tab:auroc09all} we summarize very accurately predictable statements (AUROC>0.9) aggregated at a 3-digit level. In \Cref{tab:auroc08all}-\Cref{tab:auroc08part3all} we summarize accurately predictable statements (AUROC>0.8). Finally in \Cref{tab:auroc07xall}-\Cref{tab:auroc07xpart3all}, we show challenging statements (AUROC<0.7). All statements refer to the model trained on ALL (ED and hospital) ECGs.

\begin{table}[ht]
\centering
\scalebox{0.60}{
\begin{tabular}{lp{10cm}p{10cm}}
\toprule
Block & Code, AUROC and description & Code, AUROC and description  \\  
\midrule
\midrule
I: Infectious and Parasitic Diseases & A4151: 0.834. E. coli & A40: 0.830. Streptococcal sepsis \\
\midrule
\multirow{4}{*}{II: Neoplasms (Cancer and Tumors)} & C925: 0.874. Acute myelomonocytic leukemia  & C7952: 0.852. Secondary malignant neoplasm of bone marrow  \\        
 & C34: 0.829. Malignant neoplasm of bronchus and lung & C786: 0.822. Secondary malignant neoplasm of retroperitoneum and peritoneum    \\
 & C25: 0.813. Malignant neoplasm of pancreas & C22: 0.812. Malignant neoplasm of liver and intrahepatic bile ducts \\

\midrule
\multirow{3}{*}{III: Blood, Immune, and Hematological Disorders}& \textbf{D65: 0.902. Disseminated intravascular coagulation [defibrination syndrome]} &  D684: 0.852. Acquired coagulation factor deficiency \\
 & D631: 0.834. Anemia in chronic kidney disease & D89: 0.831. Other disorders involving the immune mechanism   \\

\midrule
\multirow{5}{*}{IV: Endocrine, Nutritional, and Metabolic Diseases}& \textbf{E660: 0.913. Obesity due to excess calories}  & E1129: 0.881. Type 2 diabetes mellitus with other diabetic kidney complications \\
 & E103: 0.878. Type 1 diabetes mellitus with ophthalmic complications & E43: 0.864. Unspecified severe protein-calorie malnutrition  \\
 & E134: 0.839. Other specified diabetes mellitus with neurological complications  & E7800: 0.820. Pure hypercholesterolemia, unspecified  \\
 
\midrule
\multirow{4}{*}{V: Mental and Behavioral Disorders} & F1022: 0.881. Alcohol dependence with intoxication  & F11: 0.855. Opioid-related disorders  \\
 & F1410: 0.828. Cocaine abuse, uncomplicated  & F129: 0.826. Cannabis use, unspecified \\
 & F068: 0.813. Other specified mental disorders due to known physiological conditions & F4310: 0.812. Post-traumatic stress disorder, unspecified  \\

\midrule
\multirow{2}{*}{VI: Nervous System Disorders} & G931: 0.859. Anoxic brain damage, not elsewhere classified  &  G309: 0.807. Alzheimer's disease, unspecified  \\
 & G20: 0.803. Parkinson's disease  &   \\

\midrule
\multirow{7}{*}{IX: Circulatory System Diseases} &  \textbf{I2109: 0.983. ST elevation (STEMI) myocardial infarction involving other coronary artery of anterior wall}   & \textbf{I447: 0.959. Left bundle-branch block, unspecified}   \\
 & \textbf{I255: 0.949. Ischemic cardiomyopathy}   &  \textbf{I314: 0.947. Cardiac tamponade}  \\
 & \textbf{I4510: 0.939. Unspecified right bundle-branch block} & \textbf{I5023: 0.929. Acute on chronic systolic (congestive) heart failure}  \\
 & \textbf{I481: 0.926. Persistent atrial fibrillation} & \textbf{I428: 0.910. Other cardiomyopathies}  \\
 & \textbf{I078: 0.909. Other rheumatic tricuspid valve diseases} & \textbf{I851: 0.901. Secondary esophageal varices} \\
 & I2789: 0.897. Other specified pulmonary heart diseases & I13: 0.891. Hypertensive heart and chronic kidney disease  \\
\midrule

\multirow{4}{*}{X: Respiratory System Diseases} & J9621: 0.880. Acute and chronic respiratory failure with hypoxia  &  J910: 0.860. Malignant pleural effusion  \\
 &  J948: 0.8534. Other specified pleural conditions & J441: 0.826. Chronic obstructive pulmonary disease with (acute) exacerbation  \\
 & J47: 0.8128. Bronchiectasis & J80: 0.803. Acute respiratory distress syndrome \\

\midrule
\multirow{3}{*}{XI: Digestive System Diseases} & \textbf{K767: 0.953. Hepatorenal syndrome}  & \textbf{K701: 0.921. Alcoholic hepatitis}  \\
 & \textbf{K7291: 0.909. Hepatic failure, unspecified with coma}  & \textbf{K3189: 0.904. Other diseases of stomach and duodenum}  \\
 & K65: 0.855. Peritonitis & K830: 0.821. Cholangitis \\

\midrule
\multirow{2}{*}{XII: Skin and Subcutaneous Tissue Diseases} & L9740: 0.842. Non-pressure chronic ulcer of unspecified heel and midfoot & L8915: 0.835. Pressure ulcer of sacral region \\
\midrule

\multirow{3}{*}{XIV: Genitourinary System Diseases} & N186: 0.869. End-stage renal disease &  N08: 0.832. Glomerular disorders in diseases classified elsewhere  \\
 & N170: 0.822. Acute kidney failure with tubular necrosis & N401: 0.813. Benign prostatic hyperplasia with lower urinary tract symptoms \\

\midrule
XVII: Congenital Abnormalities & Q23: 0.866. Congenital malformations of aortic and mitral valves  &         \\
\midrule
\multirow{3}{*}{XVIII: Symptoms and Clinical Findings} & \textbf{R570: 0.916. Cardiogenic shock} & R64: 0.893. Cachexia  \\
 & R18: 0.876. Ascites  & R6521: 0.870. Severe sepsis with septic shock   \\
 &  R000: 0.831. Tachycardia, unspecified  & R402: 0.819. Coma  \\

\midrule
\multirow{3}{*}{XIX: Injuries and External Causes} & \textbf{T8612: 0.903. Kidney transplant failure}  & T811: 0.813. Postprocedural shock \\
 & T8289: 0.807. Other specified complications of cardiac and vascular prosthetic devices, implants, and grafts & T4551: 0.805. Poisoning by, adverse effect of and underdosing of anticoagulants     \\
 
\midrule
\multirow{6}{*}{XX: External Causes of Morbidity and Mortality} & \textbf{V850: 0.930. Driver of special construction vehicle injured in traffic accident} & \textbf{V422: 0.920. Person on outside of car injured in collision with two- or three-wheeled motor vehicle in nontraffic accident} \\
 & \textbf{V433: 0.904. Unspecified car occupant injured in collision with car, pick-up truck, or van in nontraffic accident}  &  V462: 0.871. Person on outside of car injured in collision with other nonmotor vehicle in nontraffic accident  \\
 & V600: 0.865. Driver of heavy transport vehicle injured in collision with pedestrian or animal in nontraffic accident & V667: 0.848. Person on outside of heavy transport vehicle injured in collision with other nonmotor vehicle in traffic accident \\

\midrule
\multirow{4}{*}{XXI: Health Status and Services} & \textbf{Z4502: 0.960. Encounter for adjustment and management of automatic implantable cardiac defibrillator} &  \textbf{Z9581: 0.935. Presence of other cardiac implants and grafts}   \\
 & \textbf{Z681: 0.913. Body mass index (BMI) 19.9 or less, adult} & \textbf{Z9981: 0.912. Dependence on supplemental oxygen}  \\
 & Z7682: 0.888. Awaiting organ transplant status & Z590: 0.882. Homelessness  \\

\bottomrule
\end{tabular}
}

\caption{T(ALL2ALL)-E(ALL2ALL) model: Best-performing individual statements organized according to ICD chapters. The table shows the six best-performing individual statements per ICD chapter (10 for chapter IX), where we show only AUROC scores above 0.8, see also \Cref{tab:auroc09all}, \Cref{tab:auroc08all}-\Cref{tab:auroc08part3all}, and  \Cref{tab:auroc07xall}-\Cref{tab:auroc07xpart3all} for a summary corresponding summary of ICD codes at 3-digit level with AUROC scores above 0.9, 0.8 and below 0.7, respectively. To showcase the breadth of reliably predictable statements by listing only the best-performing statement per 3-digit ICD code. Statements with AUROC score of 0.9 or higher are marked in boldface.}

\label{tab:big_all_ALL}
\end{table}

\begin{table}[ht]
    \centering
    \scalebox{0.8}{
    \begin{tabular}{llllp{8cm}}
    \toprule
Chapter & Code & Coverage & Prevalence & Description\\
\midrule
III & \textbf{D65} & 1/1 &0.002 & Disseminated intravascular coagulation [defibrination syndrome]\\
\midrule
IV & E66 & 2/4 &0.081 & Overweight and obesity\\
\midrule
IX & I25 & 2/10 &0.229 & Chronic ischemic heart disease\\
IX & I48 & 3/7 &0.204 & Atrial fibrillation and flutter\\
IX & I50 & 4/15 &0.194 & Heart failure\\
IX & I21 & 4/8 &0.054 & Acute myocardial infarction\\
IX & I42 & 1/3 &0.028 & Cardiomyopathy\\
IX & \textbf{I44} & 4/5 &0.024 & Atrioventricular and left bundle-branch block\\
IX & I45 & 2/5 &0.015 & Other conduction disorders\\
IX & I31 & 1/4 &0.011 & Other diseases of pericardium\\
IX & I07 & 1/2 &0.009 & Rheumatic tricuspid valve diseases\\
IX & I85 & 1/3 &0.009 & Esophageal varices\\
\midrule
XI & K76 & 1/6 &0.032 & Other diseases of liver\\
XI & K31 & 1/4 &0.027 & Other diseases of stomach and duodenum\\
XI & K70 & 2/5 &0.019 & Alcoholic liver disease\\
XI & K72 & 1/6 &0.009 & Hepatic failure, not elsewhere classified\\
\midrule
XVIII & R57 & 1/4 &0.015 & Shock, not elsewhere classified\\
\midrule
XIX & T86 & 1/3 &0.010 & Complications of transplanted organs and tissue\\
\midrule
XX & V43 & 1/4 &0.020 & Car occupant injured in collision with car, pick-up truck or van\\
XX & V42 & 1/3 &0.010 & Car occupant injured in collision with two- or three-wheeled motor vehicle\\
XX & V85 & 1/2 &0.008 & Occupant of special construction vehicle injured in transport accident\\
\midrule
XXI & Z95 & 4/7 &0.106 & Presence of cardiac and vascular implants and grafts\\
XXI & Z68 & 2/8 &0.072 & Body mass index [BMI]\\
XXI & Z99 & 2/5 &0.036 & Dependence on enabling machines and devices, not elsewhere classified\\
XXI & \textbf{Z45} & 3/4 &0.007 & Encounter for adjustment and management of implanted device\\    
\bottomrule
    \end{tabular}
    }
    \caption{T(ALL2ALL)-E(ALL2ALL) model: Statements with AUROCs larger than 0.9. The statements are sorted by chapter and prevalence in the dataset. Coverage refers to the fraction of codes within this 3-digit category that exceed the specified threshold of 0.9. Categories in boldface correspond to categories with a coverage of 75\% or higher, i.e., a situation where the category and 75\% of the subcategories are covered with AUROC scores above 0.9.}
    \label{tab:auroc09all}
\end{table}

\begin{table}[ht]
    \centering
    \scalebox{0.8}{
    \begin{tabular}{llllp{8cm}}
    \toprule
Chapter & Code & Coverage & Prevalence & Description\\
\midrule
I & A41 & 4/7 &0.049 & Other sepsis\\
I & \textbf{A40} & 1/1 &0.003 & Streptococcal sepsis\\
\midrule
II & C79 & 3/8 &0.019 & Secondary malignant neoplasm of other and unspecified sites\\
II & C78 & 2/5 &0.018 & Secondary malignant neoplasm of respiratory and digestive organs\\
II & C34 & 1/4 &0.014 & Malignant neoplasm of bronchus and lung\\
II & \textbf{C25} & 1/1 &0.005 & Malignant neoplasm of pancreas\\
II & \textbf{C92} & 2/2 &0.005 & Myeloid leukemia\\
II & \textbf{C22} & 1/1 &0.005 & Malignant neoplasm of liver and intrahepatic bile ducts\\
\midrule
III & D63 & 2/4 &0.052 & Anemia in chronic diseases classified elsewhere\\
III & D68 & 2/4 &0.032 & Other coagulation defects\\
III & \textbf{D89} & 1/1 &0.004 & Other disorders involving the immune mechanism, not elsewhere classified\\
III & \textbf{D65} & 1/1 &0.002 & Disseminated intravascular coagulation [defibrination syndrome]\\
\midrule
IV & E78 & 1/4 &0.353 & Disorders of lipoprotein metabolism and other lipidemias\\
IV & E11 & 10/18 &0.248 & Type 2 diabetes mellitus\\
IV & E87 & 1/11 &0.186 & Other disorders of fluid, electrolyte and acid-base balance\\
IV & \textbf{E66} & 3/4 &0.081 & Overweight and obesity\\
IV & \textbf{E13} & 3/3 &0.024 & Other specified diabetes mellitus\\
IV & E10 & 5/7 &0.022 & Type 1 diabetes mellitus\\
IV & \textbf{E43} & 1/1 &0.014 & Unspecified severe protein-calorie malnutrition\\
\midrule
V & F10 & 4/8 &0.061 & Alcohol related disorders\\
V & F43 & 1/5 &0.020 & Reaction to severe stress, and adjustment disorders\\
V & F02 & 1/3 &0.015 & Dementia in other diseases classified elsewhere\\
V & \textbf{F11} & 5/5 &0.014 & Opioid related disorders\\
V & \textbf{F14} & 3/3 &0.012 & Cocaine related disorders\\
V & \textbf{F12} & 2/2 &0.009 & Cannabis related disorders\\
V & F06 & 1/2 &0.006 & Other mental disorders due to known physiological condition\\
\midrule
VI & G93 & 1/7 &0.041 & Other disorders of brain\\
VI & \textbf{G30} & 2/2 &0.011 & Alzheimer's disease\\
VI & \textbf{G20} & 1/1 &0.008 & Parkinson's disease\\
\midrule
IX & \textbf{I25} & 9/10 &0.229 & Chronic ischemic heart disease\\
IX & \textbf{I48} & 7/7 &0.204 & Atrial fibrillation and flutter\\
IX & \textbf{I50} & 13/15 &0.194 & Heart failure\\
IX & I12 & 1/3 &0.124 & Hypertensive chronic kidney disease\\
IX & \textbf{I21} & 8/8 &0.054 & Acute myocardial infarction\\
IX & \textbf{I27} & 5/5 &0.050 & Other pulmonary heart diseases\\
IX & \textbf{I34} & 3/3 &0.031 & Nonrheumatic mitral valve disorders\\
IX & \textbf{I42} & 3/3 &0.028 & Cardiomyopathy\\
IX & I47 & 1/3 &0.028 & Paroxysmal tachycardia\\
IX & \textbf{I13} & 3/3 &0.027 & Hypertensive heart and chronic kidney disease\\
IX & \textbf{I35} & 3/3 &0.025 & Nonrheumatic aortic valve disorders\\
IX & \textbf{I44} & 5/5 &0.024 & Atrioventricular and left bundle-branch block\\
IX & \textbf{I11} & 2/2 &0.024 & Hypertensive heart disease\\
IX & \textbf{I70} & 4/5 &0.023 & Atherosclerosis\\
IX & I20 & 2/3 &0.020 & Angina pectoris\\
IX & I69 & 2/6 &0.020 & Sequelae of cerebrovascular disease\\
IX & I71 & 1/3 &0.020 & Aortic aneurysm and dissection\\
IX & \textbf{I45} & 5/5 &0.015 & Other conduction disorders\\
IX & I97 & 1/3 &0.014 & Intraoperative and postprocedural complications and disorders of circulatory system, not elsewhere classified\\
\bottomrule
    \end{tabular}
    }
    \caption{T(ALL2ALL)-E(ALL2ALL) model: Statements with AUROCs larger than 0.8 (Part 1). The statements are sorted by chapter and prevalence in the dataset. Coverage refers to the fraction of codes within this 3-digit category that exceed the specified threshold of 0.8. Categories in boldface correspond to categories with a coverage of 75\% or higher, i.e., a situation where the category and 75\% of the subcategories are covered with AUROC scores above 0.8.}
    \label{tab:auroc08all}
\end{table}

\begin{table}[ht]
    \centering
    \scalebox{0.8}{
    \begin{tabular}{llllp{8cm}}
    \toprule
     Chapter & Code & Coverage & Prevalence & Description\\
   \midrule
IX & \textbf{I08} & 3/3 &0.011 & Multiple valve diseases\\
IX & \textbf{I31} & 4/4 &0.011 & Other diseases of pericardium\\
IX & I65 & 1/3 &0.011 & Occlusion and stenosis of precerebral arteries, not resulting in cerebral infarction\\
IX & \textbf{I07} & 2/2 &0.009 & Rheumatic tricuspid valve diseases\\
IX & \textbf{I85} & 3/3 &0.009 & Esophageal varices\\
IX & I51 & 1/2 &0.009 & Complications and ill-defined descriptions of heart disease\\
IX & \textbf{I46} & 2/2 &0.006 & Cardiac arrest\\
IX & \textbf{I16} & 1/1 &0.003 & Hypertensive crisis\\
IX & \textbf{I33} & 2/2 &0.002 & Acute and subacute endocarditis\\
\midrule
X & J44 & 1/3 &0.088 & Other chronic obstructive pulmonary disease\\
X & \textbf{J96} & 8/10 &0.067 & Respiratory failure, not elsewhere classified\\
X & \textbf{J91} & 3/3 &0.022 & Pleural effusion in conditions classified elsewhere\\
X & J95 & 1/4 &0.015 & Intraoperative and postprocedural complications and disorders of respiratory system, not elsewhere classified\\
X & J15 & 1/4 &0.011 & Bacterial pneumonia, not elsewhere classified\\
X & \textbf{J47} & 1/1 &0.003 & Bronchiectasis\\
X & \textbf{J94} & 2/2 &0.003 & Other pleural conditions\\
X & \textbf{J80} & 1/1 &0.002 & Acute respiratory distress syndrome\\
\midrule
XI & K76 & 4/6 &0.032 & Other diseases of liver\\
XI & \textbf{K31} & 4/4 &0.027 & Other diseases of stomach and duodenum\\
XI & \textbf{K70} & 5/5 &0.019 & Alcoholic liver disease\\
XI & K83 & 1/3 &0.013 & Other diseases of biliary tract\\
XI & \textbf{K72} & 6/6 &0.009 & Hepatic failure, not elsewhere classified\\
XI & \textbf{K65} & 2/2 &0.007 & Peritonitis\\
XI & K66 & 1/2 &0.005 & Other disorders of peritoneum\\
\midrule
XII & \textbf{L89} & 3/3 &0.020 & Pressure ulcer\\
XII & \textbf{L97} & 4/5 &0.020 & Non-pressure chronic ulcer of lower limb, not elsewhere classified\\
\midrule
XIV & N18 & 2/7 &0.171 & Chronic kidney disease (CKD)\\
XIV & N17 & 1/3 &0.165 & Acute kidney failure\\
XIV & \textbf{N40} & 3/3 &0.057 & Benign prostatic hyperplasia\\
XIV & \textbf{N08} & 1/1 &0.008 & Glomerular disorders in diseases classified elsewhere\\
\midrule
XVII & \textbf{Q23} & 1/1 &0.003 & Congenital malformations of aortic and mitral valves\\
\midrule
XVIII & R00 & 1/4 &0.080 & Abnormalities of heart beat\\
XVIII & R06 & 1/6 &0.039 & Abnormalities of breathing\\
XVIII & \textbf{R65} & 4/4 &0.036 & Symptoms and signs specifically associated with systemic inflammation and infection\\
XVIII & R50 & 1/4 &0.025 & Fever of other and unknown origin\\
XVIII & \textbf{R57} & 3/4 &0.015 & Shock, not elsewhere classified\\
XVIII & \textbf{R18} & 2/2 &0.015 & Ascites\\
XVIII & R40 & 1/3 &0.011 & Somnolence, stupor and coma\\
XVIII & \textbf{R64} & 1/1 &0.009 & Cachexia\\
XVIII & \textbf{R34} & 1/1 &0.003 & Anuria and oliguria\\
\midrule
XIX & T82 & 1/7 &0.024 & Complications of cardiac and vascular prosthetic devices, implants and grafts\\
XIX & T81 & 1/5 &0.014 & Complications of procedures, not elsewhere classified\\
XIX & T45 & 1/4 &0.013 & Poisoning by, adverse effect of and underdosing of primarily systemic and hematological agents, not elsewhere classified\\
XIX & \textbf{T86} & 3/3 &0.010 & Complications of transplanted organs and tissue\\
    \bottomrule
    \end{tabular}
    }
    \caption{T(ALL2ALL)-E(ALL2ALL) model: Statements with AUROCs larger than 0.8 (Part 2). The statements are sorted by chapter and prevalence in the dataset. Coverage refers to the fraction of codes within this 3-digit category that exceed the specified threshold of 0.8. Categories in boldface correspond to categories with a coverage of 75\% or higher, i.e., a situation where the category and 75\% of the subcategories are covered with AUROC scores above 0.8.}
    \label{tab:auroc08part2all}
\end{table}

\begin{table}[ht]
    \centering
    \scalebox{0.8}{
    \begin{tabular}{llllp{8cm}}
    \toprule
    Chapter & Code & Coverage & Prevalence & Description\\
\midrule
XX & Y83 & 1/6 &0.064 & Surgical operation and other surgical procedures as the cause of abnormal reaction of the patient, or of later complication, without mention of misadventure at the time of the procedure\\
XX & Y84 & 1/4 &0.039 & Other medical procedures as the cause of abnormal reaction of the patient, or of later complication, without mention of misadventure at the time of the procedure\\
XX & V43 & 2/4 &0.020 & Car occupant injured in collision with car, pick-up truck or van\\
XX & \textbf{V10} & 2/2 &0.016 & Pedal cycle rider injured in collision with pedestrian or animal\\
XX & \textbf{V46} & 2/2 &0.012 & Car occupant injured in collision with other nonmotor vehicle\\
XX & \textbf{V42} & 3/3 &0.010 & Car occupant injured in collision with two- or three-wheeled motor vehicle\\
XX & \textbf{V85} & 2/2 &0.008 & Occupant of special construction vehicle injured in transport accident\\
XX & \textbf{V66} & 2/2 &0.008 & Occupant of heavy transport vehicle injured in collision with other nonmotor vehicle\\
XX & \textbf{V60} & 2/2 &0.006 & Occupant of heavy transport vehicle injured in collision with pedestrian or animal\\
\midrule
XXI & Z79 & 1/11 &0.265 & Long term (current) drug therapy\\
XXI & Z85 & 1/15 &0.119 & Personal history of malignant neoplasm\\
XXI & \textbf{Z95} & 7/7 &0.106 & Presence of cardiac and vascular implants and grafts\\
XXI & Z68 & 5/8 &0.072 & Body mass index [BMI]\\
XXI & Z98 & 2/6 &0.069 & Other postprocedural states\\
XXI & \textbf{Z66} & 1/1 &0.067 & Do not resuscitate\\
XXI & \textbf{Z99} & 4/5 &0.036 & Dependence on enabling machines and devices, not elsewhere classified\\
XXI & Z51 & 1/2 &0.015 & Encounter for other aftercare and medical care\\
XXI & \textbf{Z89} & 5/5 &0.014 & Acquired absence of limb\\
XXI & \textbf{Z45} & 4/4 &0.007 & Encounter for adjustment and management of implanted device\\
XXI & \textbf{Z76} & 3/3 &0.006 & Persons encountering health services in other circumstances\\
XXI & \textbf{Z59} & 2/2 &0.004 & Problems related to housing and economic circumstances\\
    \bottomrule
    \end{tabular}
    }
    \caption{T(ALL2ALL)-E(ALL2ALL) model: Statements with AUROCs larger than 0.8 (Part 3). The statements are sorted by chapter and prevalence in the dataset. Coverage refers to the fraction of codes within this 3-digit category that exceed the specified threshold of 0.8. Categories in boldface correspond to categories with a coverage of 75\% or higher, i.e., a situation where the category and 75\% of the subcategories are covered with AUROC scores above 0.8.}
    \label{tab:auroc08part3all}
\end{table}

\begin{table}[ht]
    \centering
    \scalebox{0.8}{
    \begin{tabular}{llllp{8cm}}
    \toprule
    Chapter & Code & Coverage & Prevalence & Description\\
\midrule
I & B96 & 2/9 &0.045 & Other bacterial agents as the cause of diseases classified elsewhere\\
I & B95 & 2/6 &0.030 & Streptococcus, Staphylococcus, and Enterococcus as the cause of diseases classified elsewhere\\
I & \textbf{A08} & 1/1 &0.005 & Viral and other specified intestinal infections\\
I & \textbf{B00} & 1/1 &0.001 & Herpesviral [herpes simplex] infections\\
\midrule
II & D47 & 2/3 &0.011 & Other neoplasms of uncertain behavior of lymphoid, hematopoietic and related tissue\\
II & \textbf{C91} & 3/3 &0.009 & Lymphoid leukemia\\
\midrule
III & D64 & 2/3 &0.096 & Other anemias\\
III & \textbf{D72} & 5/5 &0.040 & Other disorders of white blood cells\\
III & \textbf{D50} & 3/3 &0.036 & Iron deficiency anemia\\
III & D68 & 1/4 &0.032 & Other coagulation defects\\
III & \textbf{D86} & 1/1 &0.003 & Sarcoidosis\\
III & \textbf{D75} & 1/1 &0.002 & Other and unspecified diseases of blood and blood-forming organs\\
\midrule
IV & E78 & 1/4 &0.353 & Disorders of lipoprotein metabolism and other lipidemias\\
IV & \textbf{E03} & 2/2 &0.106 & Other hypothyroidism\\
IV & \textbf{E86} & 3/4 &0.066 & Volume depletion\\
IV & E83 & 1/9 &0.036 & Disorders of mineral metabolism\\
IV & \textbf{E55} & 2/2 &0.013 & Vitamin D deficiency\\
IV & \textbf{E27} & 3/3 &0.011 & Other disorders of adrenal gland\\
IV & \textbf{E53} & 2/2 &0.009 & Deficiency of other B group vitamins\\
IV & \textbf{E04} & 2/2 &0.008 & Other nontoxic goiter\\
IV & \textbf{E89} & 2/2 &0.008 & Postprocedural endocrine and metabolic complications and disorders, not elsewhere classified\\
IV & \textbf{E22} & 2/2 &0.008 & Hyperfunction of pituitary gland\\
IV & \textbf{E05} & 1/1 &0.007 & Thyrotoxicosis [hyperthyroidism]\\
IV & \textbf{E21} & 2/2 &0.005 & Hyperparathyroidism and other disorders of parathyroid gland\\
\midrule
V & \textbf{F32} & 2/2 &0.129 & Major depressive disorder, single episode\\
V & \textbf{F41} & 3/3 &0.098 & Other anxiety disorders\\
V & \textbf{F31} & 2/2 &0.026 & Bipolar disorder\\
V & F43 & 1/5 &0.020 & Reaction to severe stress, and adjustment disorders\\
V & \textbf{F25} & 2/2 &0.007 & Schizoaffective disorders\\
\midrule
VI & G47 & 2/6 &0.102 & Sleep disorders\\
VI & \textbf{G89} & 4/5 &0.052 & Pain, not elsewhere classified\\
VI & \textbf{G40} & 3/3 &0.032 & Epilepsy and recurrent seizures\\
VI & \textbf{G25} & 3/3 &0.012 & Other extrapyramidal and movement disorders\\
VI & \textbf{G81} & 2/2 &0.008 & Hemiplegia and hemiparesis\\
VI & \textbf{G91} & 1/1 &0.005 & Hydrocephalus\\
\midrule
VII & \textbf{H54} & 2/2 &0.009 & Blindness and low vision\\
VII & \textbf{H53} & 1/1 &0.009 & Visual disturbances\\
VII & \textbf{H26} & 2/2 &0.004 & Other cataract\\
\midrule
VIII & \textbf{H81} & 1/1 &0.005 & Disorders of vestibular function\\
\midrule
IX & I95 & 3/7 &0.075 & Hypotension\\
IX & I67 & 2/3 &0.012 & Other cerebrovascular diseases\\
IX & \textbf{I77} & 1/1 &0.007 & Other disorders of arteries and arterioles\\
IX & \textbf{I80} & 2/2 &0.002 & Phlebitis and thrombophlebitis\\
\midrule
X & J45 & 2/4 &0.076 & Asthma\\
X & J98 & 1/7 &0.033 & Other respiratory disorders\\
X & \textbf{J38} & 1/1 &0.003 & Diseases of vocal cords and larynx, not elsewhere classified\\
    \bottomrule
    \end{tabular}
    }
    \caption{T(ALL2ALL)-E(ALL2ALL) model: Statements with AUROCs smaller than 0.7 (Part 1). The statements are sorted by chapter and prevalence in the dataset. Coverage refers to the fraction of codes within this 3-digit category that stay below the specified threshold of 0.7. Categories in boldface correspond to categories with a coverage of 75\% or higher, i.e., a situation where the category and 75\% of the subcategories are covered with AUROC scores below 0.7.}
    \label{tab:auroc07xall}
\end{table}

\begin{table}[ht]
    \centering
    \scalebox{0.8}{
    \begin{tabular}{llllp{8cm}}
    \toprule
    Chapter & Code & Coverage & Prevalence & Description\\
\midrule
XI & \textbf{K21} & 2/2 &0.176 & Gastro-esophageal reflux disease\\
XI & \textbf{K59} & 4/4 &0.050 & Other functional intestinal disorders\\
XI & K92 & 1/4 &0.024 & Other diseases of digestive system\\
XI & \textbf{K57} & 3/3 &0.021 & Diverticular disease of intestine\\
XI & K56 & 1/4 &0.020 & Paralytic ileus and intestinal obstruction without hernia\\
XI & \textbf{K22} & 3/3 &0.020 & Other diseases of esophagus\\
XI & \textbf{K29} & 3/3 &0.016 & Gastritis and duodenitis\\
XI & \textbf{K80} & 3/3 &0.013 & Cholelithiasis\\
XI & \textbf{K86} & 2/2 &0.013 & Other diseases of pancreas\\
XI & \textbf{K58} & 2/2 &0.011 & Irritable bowel syndrome\\
XI & \textbf{K85} & 2/2 &0.011 & Acute pancreatitis\\
XI & K63 & 1/2 &0.009 & Other diseases of intestine\\
XI & \textbf{K64} & 2/2 &0.009 & Hemorrhoids and perianal venous thrombosis\\
XI & \textbf{K52} & 2/2 &0.009 & Other and unspecified noninfective gastroenteritis and colitis\\
XI & \textbf{K50} & 3/3 &0.009 & Crohn's disease [regional enteritis]\\
XI & \textbf{K62} & 1/1 &0.007 & Other diseases of anus and rectum\\
XI & \textbf{K25} & 1/1 &0.005 & Gastric ulcer\\
XI & \textbf{K51} & 1/1 &0.004 & Ulcerative colitis\\
XI & \textbf{K81} & 1/1 &0.003 & Cholecystitis\\
\midrule
XII & L03 & 1/5 &0.024 & Cellulitis and acute lymphangitis\\
XII & \textbf{L40} & 1/1 &0.008 & Psoriasis\\
XII & L27 & 1/2 &0.006 & Dermatitis due to substances taken internally\\
\midrule
XIII & \textbf{M54} & 4/4 &0.048 & Dorsalgia\\
XIII & \textbf{M79} & 5/6 &0.035 & Other and unspecified soft tissue disorders, not elsewhere classified\\
XIII & \textbf{M25} & 5/5 &0.021 & Other joint disorder, not elsewhere classified\\
XIII & M48 & 2/4 &0.016 & Other spondylopathies\\
XIII & \textbf{M06} & 2/2 &0.015 & Other rheumatoid arthritis\\
XIII & \textbf{M47} & 3/3 &0.013 & Spondylosis\\
XIII & M94 & 1/2 &0.013 & Other disorders of cartilage\\
XIII & M62 & 2/3 &0.012 & Other disorders of muscle\\
XIII & M17 & 1/2 &0.008 & Osteoarthritis of knee\\
XIII & \textbf{M51} & 1/1 &0.006 & Thoracic, thoracolumbar, and lumbosacral intervertebral disc disorders\\
XIII & M85 & 1/2 &0.004 & Other disorders of bone density and structure\\
\midrule
XIV & \textbf{N28} & 2/2 &0.009 & Other disorders of kidney and ureter, not elsewhere classified\\
XIV & \textbf{N32} & 1/1 &0.005 & Other disorders of bladder\\
XIV & \textbf{N20} & 2/2 &0.004 & Calculus of kidney and ureter\\
\midrule
XVII & \textbf{Q21} & 2/2 &0.004 & Congenital malformations of cardiac septa\\
\midrule
XVIII & R00 & 1/4 &0.080 & Abnormalities of heart beat\\
XVIII & R10 & 5/7 &0.050 & Abdominal and pelvic pain\\
XVIII & \textbf{R11} & 4/4 &0.041 & Nausea and vomiting\\
XVIII & \textbf{R06} & 5/6 &0.039 & Abnormalities of breathing\\
XVIII & \textbf{R55} & 1/1 &0.039 & Syncope and collapse\\
XVIII & \textbf{R19} & 2/2 &0.032 & Other symptoms and signs involving the digestive system and abdomen\\
XVIII & R33 & 2/3 &0.031 & Retention of urine\\
XVIII & R79 & 2/4 &0.026 & Other abnormal findings of blood chemistry\\
XVIII & R50 & 1/4 &0.025 & Fever of other and unknown origin\\
XVIII & \textbf{R51} & 1/1 &0.024 & Headache\\
XVIII & R13 & 1/4 &0.024 & Aphagia and dysphagia\\
XVIII & \textbf{R53} & 3/4 &0.023 & Malaise and fatigue\\
XVIII & \textbf{R41} & 3/4 &0.022 & Other symptoms and signs involving cognitive functions and awareness\\
XVIII & R45 & 1/4 &0.021 & Symptoms and signs involving emotional state\\
XVIII & \textbf{R74} & 3/3 &0.021 & Abnormal serum enzyme levels\\
XVIII & \textbf{R91} & 3/3 &0.017 & Abnormal findings on diagnostic imaging of lung\\
XVIII & \textbf{R47} & 4/4 &0.017 & Speech disturbances, not elsewhere classified\\
XVIII & \textbf{R29} & 4/4 &0.016 & Other symptoms and signs involving the nervous and musculoskeletal systems\\
    \bottomrule
    \end{tabular}
    }
    \caption{T(ALL2ALL)-E(ALL2ALL) model: Statements with AUROCs smaller than 0.7 (Part 2). The statements are sorted by chapter and prevalence in the dataset. Coverage refers to the fraction of codes within this 3-digit category that stay below the specified threshold of 0.7. Categories in boldface correspond to categories with a coverage of 75\% or higher, i.e., a situation where the category and 75\% of the subcategories are covered with AUROC scores below 0.7.}
    \label{tab:auroc07xpart2all}
\end{table}

\begin{table}[ht]
    \centering
    \scalebox{0.8}{
    \begin{tabular}{llllp{8cm}}
    \toprule
    Chapter & Code & Coverage & Prevalence & Description\\
\midrule
XVIII & R31 & 2/3 &0.015 & Hematuria\\
XVIII & \textbf{R05} & 1/1 &0.014 & Cough\\
XVIII & R60 & 1/2 &0.014 & Edema, not elsewhere classified\\
XVIII & R63 & 1/2 &0.014 & Symptoms and signs concerning food and fluid intake\\
\midrule
XVIII & R94 & 1/3 &0.012 & Abnormal results of function studies\\
XVIII & \textbf{R26} & 2/2 &0.012 & Abnormalities of gait and mobility\\
XVIII & R40 & 1/3 &0.011 & Somnolence, stupor and coma\\
XVIII & \textbf{R56} & 2/2 &0.011 & Convulsions, not elsewhere classified\\
XVIII & R20 & 1/2 &0.011 & Disturbances of skin sensation\\
XVIII & R04 & 1/2 &0.010 & Hemorrhage from respiratory passages\\
XVIII & \textbf{R73} & 3/3 &0.009 & Elevated blood glucose level\\
XVIII & \textbf{R82} & 1/1 &0.006 & Other and unspecified abnormal findings in urine\\
XVIII & \textbf{R21} & 1/1 &0.005 & Rash and other nonspecific skin eruption\\
XVIII & \textbf{R25} & 1/1 &0.005 & Abnormal involuntary movements\\
XVIII & \textbf{R68} & 1/1 &0.004 & Other general symptoms and signs\\
XVIII & \textbf{R59} & 1/1 &0.004 & Enlarged lymph nodes\\
\midrule
XIX & T82 & 2/7 &0.024 & Complications of cardiac and vascular prosthetic devices, implants and grafts\\
XIX & S06 & 2/3 &0.017 & Intracranial injury\\
XIX & \textbf{S22} & 2/2 &0.011 & Fracture of rib(s), sternum and thoracic spine\\
XIX & \textbf{S01} & 1/1 &0.009 & Open wound of head\\
XIX & \textbf{T40} & 2/2 &0.008 & Poisoning by, adverse effect of and underdosing of narcotics and psychodysleptics [hallucinogens]\\
XIX & \textbf{S32} & 1/1 &0.006 & Fracture of lumbar spine and pelvis\\
XIX & \textbf{S42} & 1/1 &0.006 & Fracture of shoulder and upper arm\\
XIX & \textbf{S82} & 1/1 &0.005 & Fracture of lower leg, including ankle\\
XIX & T88 & 1/2 &0.004 & Other complications of surgical and medical care, not elsewhere classified\\
\midrule
XX & Y92 & 7/12 &0.138 & Place of occurrence of the external cause\\
XX & Y83 & 1/6 &0.064 & Surgical operation and other surgical procedures as the cause of abnormal reaction of the patient, or of later complication, without mention of misadventure at the time of the procedure\\
XX & Y84 & 1/4 &0.039 & Other medical procedures as the cause of abnormal reaction of the patient, or of later complication, without mention of misadventure at the time of the procedure\\
XX & W18 & 1/5 &0.027 & Other slipping, tripping and stumbling and falls\\
XX & \textbf{X58} & 1/1 &0.010 & Exposure to other specified factors\\
XX & \textbf{V64} & 1/1 &0.002 & Occupant of heavy transport vehicle injured in collision with heavy transport vehicle or bus\\
\midrule
XXI & Z79 & 2/11 &0.265 & Long term (current) drug therapy\\
XXI & \textbf{Z87} & 7/7 &0.189 & Personal history of other diseases and conditions\\
XXI & Z85 & 5/15 &0.119 & Personal history of malignant neoplasm\\
XXI & \textbf{Z86} & 8/10 &0.116 & Personal history of certain other diseases\\
XXI & Z98 & 2/6 &0.069 & Other postprocedural states\\
XXI & Z91 & 2/6 &0.053 & Personal risk factors, not elsewhere classified\\
XXI & Z90 & 5/9 &0.034 & Acquired absence of organs, not elsewhere classified\\
XXI & \textbf{Z23} & 1/1 &0.027 & Encounter for immunization\\
XXI & Z82 & 1/3 &0.017 & Family history of certain disabilities and chronic diseases (leading to disablement)\\
XXI & \textbf{Z80} & 1/1 &0.008 & Family history of primary malignant neoplasm\\    
\bottomrule
    \end{tabular}
    }
    \caption{T(ALL2ALL)-E(ALL2ALL) model: Statements with AUROCs smaller than 0.7 (Part 3). The statements are sorted by chapter and prevalence in the dataset. Coverage refers to the fraction of codes within this 3-digit category that stay below the specified threshold of 0.7. Categories in boldface correspond to categories with a coverage of 75\% or higher, i.e., a situation where the category and 75\% of the subcategories are covered with AUROC scores below 0.7.}
    \label{tab:auroc07xpart3all}
\end{table}

\bibliographySM{bibfile}
\bibliographystyleSM{ieeetr}

\end{document}